# ASb$_3$Mn$_9$O$_{19}$ (A = K or Rb): New Mn-Based Two-Dimensional Magnetoplumbites with Geometric and Magnetic Frustration

*Jianyi Chen, Stuart Calder, Joseph A. M. Paddison, Gina Angelo, Liana Klivansky, Jian Zhang, Huibo Cao and Xin Gui\**

Jianyi Chen, Gina Angelo, Xin Gui
Department of Chemistry, University of Pittsburgh, Pittsburgh, PA, 15260, USA
E-mail: xig75@pitt.edu
Stuart Calder, Joseph A. M. Paddison, Huibo Cao
Neutron Scattering Division, Oak Ridge National Laboratory, Oak Ridge, TN, 37831, USA
Liana Klivansky, Jian Zhang
The Molecular Foundry, Lawrence Berkeley National Laboratory, Berkeley, CA, 94720, USA

**Keywords:** magnetoplumbites, Kagome lattice, magnetic frustration, 2-D magnetic correlation

## *Abstract*

Magnetoplumbites are one of the most broadly studied families of hexagonal ferrites, typically with high magnetic ordering temperatures, making them excellent candidates for permanent magnets. However, magnetic frustration was rarely observed in magnetoplumbites. Herein, we report the discovery, synthesis and characterization of the first Mn-based magnetoplumbite, as well as the first magnetoplumbite involving pnictogens (Sb), ASb$_3$Mn$_9$O$_{19}$ (A = K or Rb). The Mn$^{3+}$ ($S$ = 2) cations, further confirmed by DC magnetic susceptibility and X-ray photoelectron spectroscopy, construct three geometrically frustrated sublattices, including Kagome, triangular and puckered honeycomb lattices. Magnetic properties measurements revealed strong antiferromagnetic spin-spin coupling as well as multiple low-temperature magnetic features. Heat capacity data did not show any prominent λ-anomaly, suggesting minimal associated magnetic entropy. Moreover, neutron powder diffraction implied the absence of long-range magnetic ordering in KSb$_3$Mn$_9$O$_{19}$ down to 3 K. However, several magnetic peaks were observed in RbSb$_3$Mn$_9$O$_{19}$ at 3 K, corresponding to an incommensurate magnetic structure. Interestingly, strong diffuse scattering was seen in the neutron powder diffraction patterns of both compounds at low angles, and was analyzed by reverse Monte Carlo refinements, indicating short-range spin ordering related to frustrated magnetism as well as two-dimensional magnetic correlations in $ASb_3Mn_9O_{19}$ (A = K or Rb).

## 1. Introduction

Magnetic frustration originates from the competition between multiple magnetic exchange interactions, normally due to geometrically frustrated crystal lattices or chemical disorder.[1–7] Geometrical magnetic frustration in quantum materials has drawn tremendous attention and is of great importance in the material chemistry/physics community. Many intriguing quantum states have been proposed/observed in a variety of material systems due to geometric frustration, e.g., spin liquid/quantum spin liquid[8–14] and quantum spin ice[15–18]. Despite being investigated for decades, there still exists a long-standing need for the discovery of new magnetically frustrated materials due to limitations in existing systems, e.g., chemical disorder that can lead to ambiguity in observing quantum spin liquid states.[1,2,10] Chemical design plays a crucial role in expanding the pool of frustrated magnets, while starting from specific magnetic crystal lattices is proven to be one of the most effective means to achieve such a goal. For instance, a variety of frustrated magnets with magnetic triangular,[9,12,13,19,20] Kagome,[21–23] honeycomb,[24–27] pyrochlore[11,16–18] and square net[28,29] lattices have been discovered and investigated. Ferrites with a spinel formula of $AB_2O_4$ and cubic symmetry, as one of the most well-known and heavily studied families of magnetic materials, were explored for their high magnetic ordering temperatures and developed to serve as permanent magnets in many applications.[30–32] Moreover, magnetic frustration has also been commonly observed in $AB_2O_4$ ferrites due to the existence of a pyrochlore lattice of B-site ions, for instance, $MCr_2O_4$[33,34] and $LiV_2O_4$.[35–37] Interestingly, when more chemical complexities are involved in $AB_2O_4$ ferrites, a higher structural/compositional tunability is induced, leaving numerous possibilities open to invent more magnetically frustrated quantum materials.

M-type hexaferrites, also known as magnetoplumbites, are one of the most widely studied subgroups of $AB_2O_4$ ferrites, adopting a general formula of $AB_{12}O_{19}$.[38,39] Here, A is mainly alkali,[40,41] alkaline-earth elements,[42,43] lanthanides,[44] Pb[45] or a mixture of them,[46,47] while B can be group 13 elements,[48–50] transition metal elements including Ti, V, Cr, Fe, Co and Ni,[45,47,51,52] or a mixture.[53–56] They typically crystallize in a hexagonal unit cell with a space group of $P6_3/mmc$, where A-site ions are well separated by polyhedra formed by B and O and the cation B occupies various atomic sites. In terms of magnetic properties, magnetoplumbites with magnetic B cations are usually considered as great candidates for permanent magnets due to their high Curie temperatures.[39] Interestingly, several sublattices of B can be found in $AB_{12}O_{19}$, e.g., triangular, Kagome and puckered honeycomb sublattices, which makes magnetoplumbites a promising material platform for inducing frustrated magnetism. However, only very limited examples of magnetoplumbites have been reported to show magnetic frustration, including spin glass in $MCr_{9p}Ga_{12-9p}O_{19}$ (M = Sr, Ba),[57–64] two-dimensional magnetic frustration in $LnMgAl_{11}O_{19}$ (Ln = Pr, Nd) and $LnZnAl_{11}O_{19}$ (Ln = Pr, Nd, Sm, Eu, Gd, Tb),[65,66] spin-glass state in $SrCo_6Ti_6O_{19}$[67,68] and $BaFe_{12}O_{19}$,[69] as well as a large frustration factor of ~26 observed in $BaSn_6Co_6O_{19}$.[70]

Here, we present the discovery and characterization of a novel type of magnetoplumbite, $ASb_3Mn_9O_{19}$ (A = K or Rb). To the best of our knowledge, they are the first Mn-based magnetoplumbites, as well as the first magnetoplumbites involving pnictogens (Sb). Polycrystalline samples were synthesized and characterized, and they both adopt a magnetoplumbite structure. According to the single crystal X-ray diffraction, we determined that there are three distinct Mn sites in $ASb_3Mn_9O_{19}$, forming a Kagome, a puckered honeycomb, and a triangular sublattice, respectively. The magnetic properties and heat capacity measurements reveal several low-temperature magnetic features down to 1.8 K. The Curie-Weiss fitting on the DC magnetic susceptibility shows strong antiferromagnetic coupling between $Mn^{3+}$ ($S = 2$), while the single valency and trivalent nature of Mn are consistent with the X-ray photoelectron spectroscopy results. Neutron powder diffraction further confirms the absence of long-range ordering in $KSb_3Mn_9O_{19}$ but indicates possible incommensurate magnetic ordering of $RbSb_3Mn_9O_{19}$. We also observed strong diffuse scattering in neutron powder diffraction patterns in both $KSb_3Mn_9O_{19}$ and $RbSb_3Mn_9O_{19}$, which likely originates from frustrated magnetism as well as two-dimensional magnetic correlations. The discovery of the new insulating $ASb_3Mn_9O_{19}$, as the first Mn-based magnetoplumbite, provides a great platform for investigating frustrated magnetism in the puckered honeycomb, Kagome and triangular sublattices, as well as the intertwining properties among them. Additionally, it allows further modification of the magnetic sites, suggesting the potential for discovering more exotic quantum states, such as new integer-spin-frustrated magnets.[71–74] New quantum spin liquids may also be realized in this sytem if a $S = ½$ spin state can be achieved.

## *2. Results and Discussion*

**2.1. Crystal Structure of ASb$_3$Mn$_9$O$_{19}$:** Crystal structures of ASb$_3$Mn$_9$O$_{19}$ (A = K and Rb) are found to be similar, both of which crystallize in a hexagonal space group *P*6$_3$/*mmc* (No. 194). The crystallographic data including refined anisotropic displacement parameters and equivalent isotropic thermal displacement parameters of both compounds, are summarized in Table 1 and Tables S1 and S2 in the supporting information (SI). The crystal structure of ASb$_3$Mn$_9$O$_{19}$ is similar to the well-known structural family, M-type hexaferrite, or magnetoplumbite. Therefore, ASb$_3$Mn$_9$O$_{19}$ becomes the first example of a Mn-based magnetoplumbite as well as the first magnetoplumbite involving pnictogens (Sb). The crystal structure of $ASb_3Mn_9O_{19}$ is shown in Figure 1a while the coordination of metal cations is illustrated in Figures 1b and 1c. As seen in Figure 1b, three crystallographically distinct Mn sites can be found in $ASb_3Mn_9O_{19}$, namely Mn1 (Wyck. 4$f$), Mn2 (Wyck. 12$k$) and Mn3 (Wyck. 2$b$) sites. Each Mn site coordinates differently with oxygen atoms, i.e., Mn1, Mn2 and Mn3 occupy the center of Mn1@$O_4$ tetrahedron, Mn2@$O_6$ octahedron and Mn3@$O_5$ trigonal bipyramid, respectively. Meanwhile, Sb atoms occupy two crystallographic sites, Sb1 (Wyck. 4$f$) and Sb2 (Wyck. 2$a$) sites, and A possesses only one atomic site (Wyck. 2$d$). Both Sb sites are 6-coordinated with oxygen, building Sb@$O_6$ octahedra, while A coordinates with 12 oxygen atoms, constructing A@$O_{12}$ cuboctahedra, as shown in Figure 1c. Moreover, Mn atoms show an interesting framework, as can be seen in Figures 1d-1f. Each crystallographically different Mn site forms a distinct Mn sublattice, Mn1 puckered honeycomb, Mn2 Kagome and Mn3 triangular sublattices. The separation between each Mn sheet is also shown in Figures 1d-1f. A well-separated puckered honeycomb sublattice of Mn1 and a well-separated triangular Mn3 sublattice can be found. The Kagome lattice of Mn2 can be treated as a "quasi-bilayer Kagome" with separation of ~5 Å while each "bilayer" is moderately separated by ~6.5 Å for both materials. Additionally, it is noteworthy that the Mn2-Mn2 bond length within the Kagome sublattice is the shortest Mn-Mn distance (3.030 (1) Å for K and 3.041 (1) Å for Rb) in both compounds. Therefore, one can expect that the primary magnetic correlations will take place within the Kagome layers, as further detailed later in the DC magnetic and neutron powder diffraction measurements.

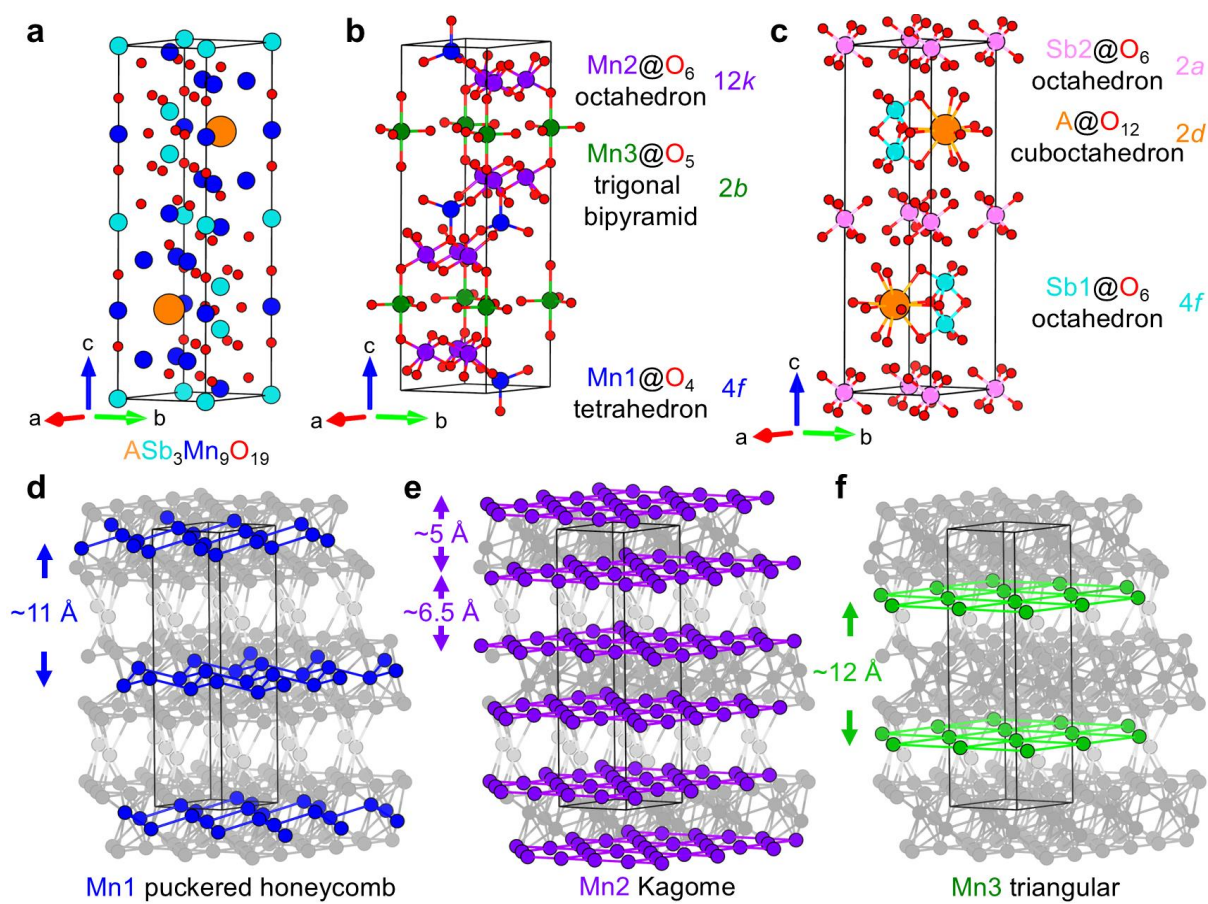

**Figure 1. a.** Crystal structure of ASb$_3$Mn$_9$O$_{19}$ (A = K or Rb) where orange, cyan, blue and red spheres represent A, Sb, Mn and O atoms, respectively. **b.** Coordination types of Mn1, Mn2 and Mn3 sites where blue, purple and green spheres stand for Mn1, Mn2 and Mn3 sites, respectively. **c.** Coordination types of A, Sb1 and Sb2 sites, represented by orange, cyan and pink spheres. **d-f.** Sublattices of Mn1, Mn2 and Mn3 sites with each site highlighted in the crystal structure as well as the separation of neighboring planes of each site.

The potential superexchange pathways are illustrated in Figure S1 in the SI. As can be seen, for Mn1-Mn1 magnetic interaction, Mn1-O-Sb2-O-Mn1 and Mn1-O-Mn2-O-Mn1 are two possible pathways due to the corner-sharing nature of Mn1@$O_4$ tetrahedron and Sb2/Mn2@$O_6$ octahedron. Because of the edge-sharing feature of Mn2@$O_6$ octahedra, the most plausible superexchange pathway is Mn2-O-Mn2. Furthermore, Mn3-Mn3 magnetic interactions are mediated by Mn3-O-Sb1-O-Mn3 pathway, due to the corner-sharing Mn3@$O_5$ trigonal bipyramid and Sb1@$O_6$ octahedron. Due to the fact that Mn1@$O_4$, Mn2@$O_6$ and Mn3@$O_5$ polyhedra are mutually corner-sharing, potential interlayer magnetic exchange interaction can also be proposed through a Mn1-O-Mn2-O-Mn3 superexchange pathway.

**Table 1.** Single crystal structure refinement for $ASb_3Mn_9O_{19}$ (A = K or Rb).

| Refined Formula | $KSb_{2.82(1)}Mn_{9.18(1)}O_{19}$ | $RbSb_{2.93(1)}Mn_9O_{19}$ |
|---|---|---|
| Temperature (K) | 293 (2) | 293 (2) |
| F.W. (g/mol) | 1190.45 | 1240.66 |
| Space group; Z | $P6_3/mmc$; 2 | $P6_3/mmc$; 2 |
| $a$(Å) | 6.0606 (1) | 6.0818 (7) |
| $c$(Å) | 23.853 (1) | 23.905 (4) |
| V (Å$^3$) | 758.76 (5) | 765.8 (2) |
| θ range (º) | 3.416-33.191 | 3.409-34.343 |
| No. reflections; $R_{int}$ | 21526; 0.0631 | 21395; 0.0401 |
| No. independent reflections | 620 | 681 |
| No. parameters | 32 | 34 |
| $R_1$: $\omega R_2$ ($I>2\delta(I)$) | 0.0304; 0.0657 | 0.0291; 0.0594 |
| Goodness of fit | 1.199 | 1.387 |
| Diffraction peak and hole (e$^-$/ Å$^3$) | 1.308; -1.268 | 1.174; -0.892 |

**2.2. Determination of Chemical Composition and Chemical Disorder:** Based on the single crystal XRD results, chemical disorder or vacancy can be assigned to the Sb2 site in $ASb_3Mn_9O_{19}$. If the Sb2 site was treated as a fully occupied ordered site, the crystallographic refinement results worsen and are not acceptable for both materials; see Tables S3 and S4 in the SI. When the Sb2 site occupancy is relaxed in refinement, 10 (1)% and 7 (1)% vacancies were found for $KSb_3Mn_9O_{19}$ and $RbSb_3Mn_9O_{19}$, respectively, therefore leading to formulas of $KSb_{2.90(1)}Mn_9O_{19}$ and $RbSb_{2.93(1)}Mn_9O_{19}$. Due to a smaller electron count of Mn than Sb, it is possible that partial Mn occupancy exists in the Sb2 site, resulting in formulas of $KSb_{2.82(1)}Mn_{9.18(1)}O_{19}$ and $RbSb_{2.87(1)}Mn_{9.13(1)}O_{19}$.

To determine accurate chemical compositions, SEM-EDS measurements were performed on multiple pieces of each sample. The EDS results are summarized in Tables S5 and S6 in the SI. For $KSb_3Mn_9O_{19}$, the EDS determines the composition to be $K_{1.00(3)}Sb_{2.96(6)}Mn_{9.44(7)}O_{19}$, while for Rb compound it is $Rb_{1.0(1)}Sb_{3.16(7)}Mn_{9.49(6)}O_{19}$. The oxygen contents are not included due to its low electron count, and thus cannot be accurately determined by EDS. Considering that the sample is semiconducting, as described later in Section 2.9, electrons from the incident beam can be trapped on the sample surface, leading to a change of effective energy of the primary electron beam. Therefore, the EDS compositions are within the error from the formulas determined by single crystal XRD. Moreover, the Mn/Sb ratio from EDS can help to confirm whether the Sb2 site is disordered with Mn or partially vacant. From Tables S5 and S6 in the SI, Mn/Sb ratios were found to be 3.19 (9) in $KSb_3Mn_9O_{19}$ and 3.00 (9) in $RbSb_3Mn_9O_{19}$. By comparing them to the Mn/Sb ratios determined by single crystal XRD, as shown in Table S7, we conclude that the formulas should be $KSb_{2.82(1)}Mn_{9.18(1)}O_{19}$ and $RbSb_{2.93(1)}Mn_9O_{19}$, determined by single crystal XRD, which are utilized in all the analysis below. The reason for using single-crystal-XRD formulas is the electron trapping effect in EDS for semiconducting samples, as stated before, which can lead to inaccuracy of the absolute compositions. Thus, Mn is mixed with Sb on the Sb2 site when A = K while Sb2 site is partially vacant in A = Rb in $ASb_3Mn_9O_{19}$. For clarity, we will still denote the compounds as $KSb_3Mn_9O_{19}$ and $RbSb_3Mn_9O_{19}$ in the rest of this study.

Lastly, the charge is not balanced in $ASb_3Mn_9O_{19}$ if the oxidation state of 3+ is assumed for both Sb, due to the starting material of $Sb_2O_3$, and Mn, evidenced by magnetic properties measurements and XPS results described later. There are two possible reasons for the unbalanced charge: Sb possesses two oxidation states, 3+ and 5+; there are oxygen vacancies in $ASb_3Mn_9O_{19}$ if both Sb and Mn are trivalent.

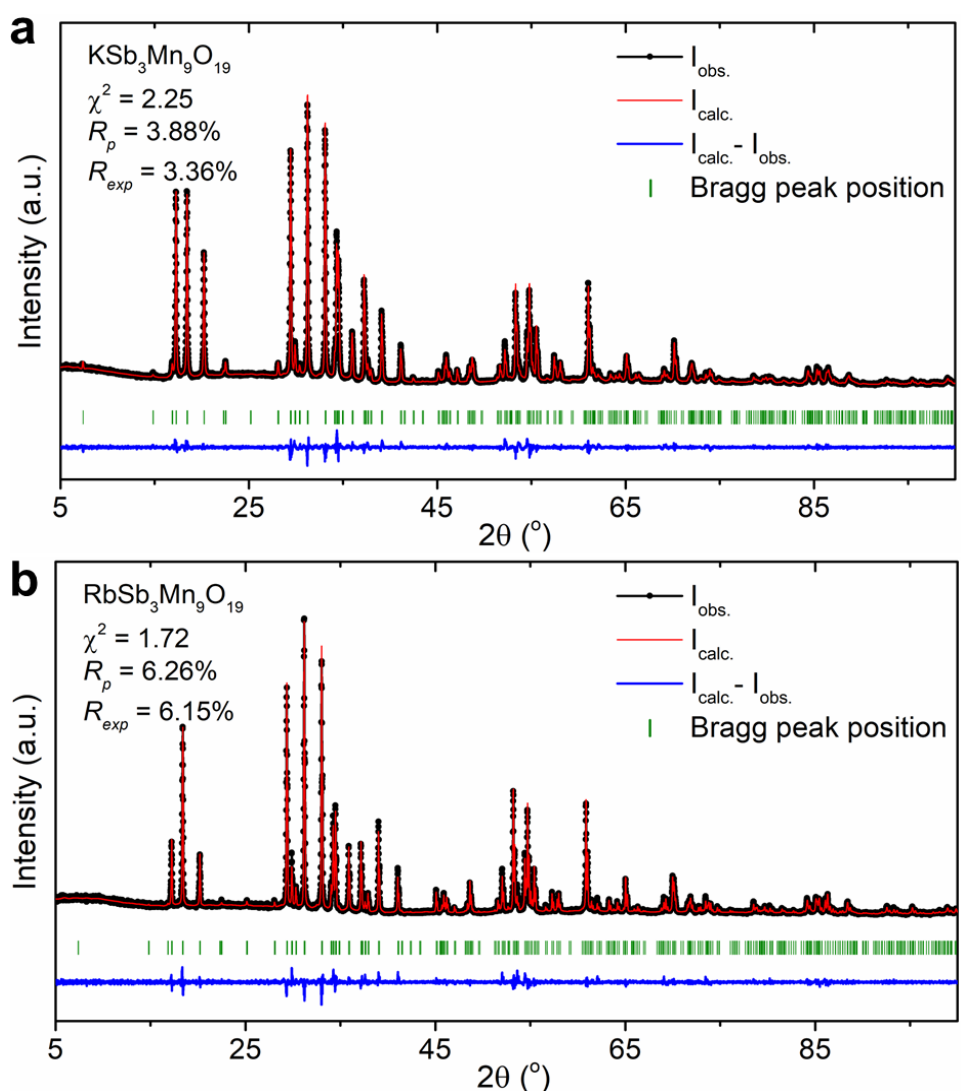

**Figure 2.** Powder XRD patterns with Rietveld refinement of **a.** $KSb_3Mn_9O_{19}$ and **b.** $RbSb_3Mn_9O_{19}$.

**2.3. Analysis of Phase Purity:** The phase purity of polycrystalline $ASb_3Mn_9O_{19}$ was determined by powder XRD via Rietveld refinement, as shown in Figures 2a and 2b. The $\chi^2$ was fit to be 2.25 and 1.72 while $R_p$ and $R_{wp}$ were found to be lower than 4% and 6.5% for $KSb_3Mn_9O_{19}$ and $RbSb_3Mn_9O_{19}$, respectively. The small difference between the observed and calculated patterns, together with the fitting parameters, imply that both materials are of high purity and that the crystal structure from single crystal XRD is consistent with powder XRD. Moreover, no prominent impurity peaks can be observed. The fitted lattice parameters are $a$ = 6.06667 (1) Å and $c$ = 23.89518 (9) Å for $KSb_3Mn_9O_{19}$, and $a$ = 6.08206 (1) Å and $c$ = 23.90777 (9) Å for $RbSb_3Mn_9O_{19}$. The slight expansion of unit cells observed in the powder XRD results, compared to those from single crystal XRD, is attributed to the elevated temperature in the sample chamber of the Bruker D2 PHASER we utilized (~40 °C).

**2.4. Magnetic Properties:** Polycrystalline samples directly from the reaction crucibles were used to measure the magnetic properties of both $KSb_3Mn_9O_{19}$ and $RbSb_3Mn_9O_{19}$. Figures 3a and 3b exhibit the temperature-dependent magnetic susceptibility ($\chi$) from 2 K to 300 K under external magnetic field of 0.3 T using zero-field-cooling (ZFC) and field-cooling (FC) protocols. For both materials, an increase at ~50 K is seen in both ZFC and FC curves while the ZFC and FC curves overlap at higher temperatures. At low temperatures, the two curves deviate at ~42 K where the FC curves keep increasing and do not reach a plateau for both samples. By plotting the temperature-dependent inverse magnetic susceptibility ($\chi^{-1}$), Curie-Weiss (CW) fitting is applied to both materials under ZFC and FC protocols from 125 K to 300 K by using the formula

$$\chi = \frac{C}{T - \theta_{CW}}$$

where C is a temperature-independent constant and is related to the effective moment ($\mu_{eff}$) via $\mu_{eff} = \sqrt{8C}$, and $\theta_{CW}$ is the CW temperature. The fitted parameters are similar between ZFC and FC modes. For $KSb_3Mn_9O_{19}$, $\theta_{CW}$ is fitted to be -152 (1) K (ZFC) and -153 (1) K (FC) while $\mu_{eff}$ is 4.946 (1) μ$_B$/Mn (ZFC) and 4.957 (1) μ$_B$/Mn (FC). For $RbSb_3Mn_9O_{19}$, significantly larger $\theta_{CW}$ and $\mu_{eff}$ are obtained. The $\theta_{CW}$ is -222 (1) K (ZFC) and -223 (1) K (FC) while $\mu_{eff}$ is 5.516 (2) μ$_B$/Mn (ZFC) and 5.508 (2) $\mu_B$/Mn (FC). According to the fitted $\theta_{CW}$, antiferromagnetic interactions can be expected within the fitted temperature range. Deviation between the CW fitting lines and the observed data can be seen at ~100 K for $KSb_3Mn_9O_{19}$ and at ~130 K for $RbSb_3Mn_9O_{19}$. Based on the fitted $\mu_{eff}$, the average oxidation state of Mn in both materials is close to $Mn^{3+}$ ($S = 2$), for which the spin-only moment is ~4.90 $\mu_B$. The larger $\mu_{eff}$ for $RbSb_3Mn_9O_{19}$ can result from unaccounted contributions from orbital angular momentum. The trivalent Mn in both compounds is also consistent with the XPS results, as detailed later in Section 2.5. Moreover, the large antiferromagnetic θ$_{CW}$ and the absence of long-range magnetic ordering above 50 K in both compounds imply the existence of magnetic frustration.

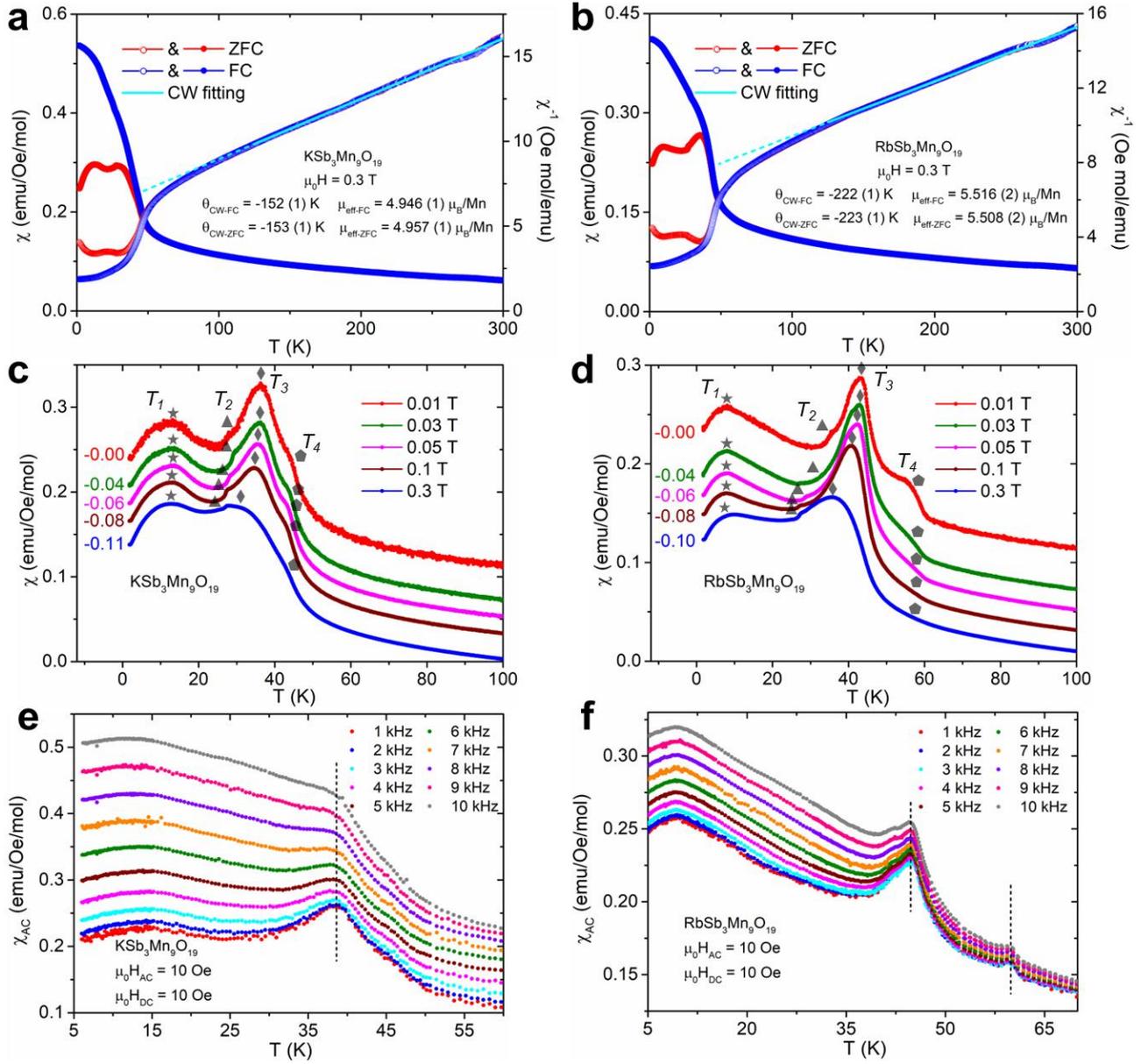

**Figure 3.** Temperature-dependent magnetic susceptibility ($\chi$) and its inverse ($\chi^{-1}$) measured under external magnetic field of 0.3 T of **a.** $KSb_3Mn_9O_{19}$ and **b.** $RbSb_3Mn_9O_{19}$. Curie-Weiss fitting curves are shown in cyan. **c.** and **d.** show the exploded view of $\chi$ under different magnetic fields for $KSb_3Mn_9O_{19}$ and $RbSb_3Mn_9O_{19}$. The curves were offset to ensure clarity. AC magnetic susceptibility ($\chi_{AC}$) measured under a DC magnetic field of 10 Oe and an AC magnetic field of 10 Oe with a variety of frequencies for **e.** $KSb_3Mn_9O_{19}$ and **f.** $RbSb_3Mn_9O_{19}$. The black dashed lines are for eye guide only.

To better investigate the low-temperature behaviors of ASb$_3$Mn$_9$O$_{19}$, Figures 3c and 3d show the ZFC curves measured under different external magnetic fields. The curves are offset vertically for clarity. Four kinks/peaks below 60 K can be seen in both compounds, denoted T$_1$, T$_2$, T$_3$ and T$_4$, and are marked using distinct shapes. The position of kinks/peaks are found to be decreasing with increasing magnetic field, consistent with antiferromagnetic interactions based on fitted $\theta_{CW}$. Considering the broad peaks corresponding to T$_1$ and T$_3$ and the deviation of ZFC and FC peaks shown in Figures 3a and 3b, the temperature-dependent AC magnetic susceptibility ($\chi_{AC}$) was measured for ASb$_3$Mn$_9$O$_{19}$. The measurements were performed under various AC frequencies ranging from 1 kHz to 10 kHz with applied DC and AC magnetic field, both of which are 10 Oe. As illustrated in Figure 3e, only two broad peaks can be observed in $\chi_{AC}$ for KSb$_3$Mn$_9$O$_{19}$, including peaks at ~14 K and ~38 K, corresponding to T$_1$ and T$_3$. However, for RbSb$_3$Mn$_9$O$_{19}$, an extra peak corresponding to T$_4$ can be found such that three groups of peaks are seen at ~9 K, ~45 K and ~60 K. None of the peaks observed in either material was found to shift towards higher temperatures with larger AC frequencies, implying that the low-temperature kinks/peaks observed in ASb$_3$Mn$_9$O$_{19}$ do not originate from conventional spin-glass state. The fact that the FC curves in both compounds do not reach a plateau below the deviation point under any applied magnetic field below 0.3 T also implies the lack of the conventional spin-glass behavior, as can be seen in Figure S2 in SI.

The hysteresis loops of ASb$_3$Mn$_9$O$_{19}$ are measured between -9 T and 9 T under different temperatures and are plotted in Figures 4a and 4b. The magnetization does not exhibit saturation at 9 T. The magnetic moments at 9 T and 1.8 K for KSb$_3$Mn$_9$O$_{19}$ ($\mu_{9T,1.8K-K}$) are determined to be $\mu_{9T,1.8K-K}$= 3.89 μ$_B$/f.u., i.e., 0.42 μ$_B$/Mn, which is only ~ 9% of the fitted $\mu_{eff}$. Similar behavior can be seen in RbSb$_3$Mn$_9$O$_{19}$ for which $\mu_{9T,1.8K-Rb}$= 3.51 μ$_B$/f.u., i.e., 0.39 μ$_B$/Mn, which is ~ 7% of the fitted $\mu_{eff}$. With increasing temperature, $\mu_{9T,1.8K}$ drops for both materials. Magnetic hysteresis is observed for both compounds under low temperatures, as shown in Figures 3c and 3d, where the hysteresis loops are shifted for clarity. The hysteresis indicates possible ferromagnetic components existing in ASb$_3$Mn$_9$O$_{19}$, which persists until 40 (45) K for KSb$_3$Mn$_9$O$_{19}$ (RbSb$_3$Mn$_9$O$_{19}$). The coercive field of KSb$_3$Mn$_9$O$_{19}$ (RbSb$_3$Mn$_9$O$_{19}$) at 1.8 K is ~0.34 T (0.06 T) while the remanent moment is ~0.22 µ$_B$/f.u. (~0.14 µ$_B$/f.u.). A slight difference of the hysteresis loops can be observed in ASb$_3$Mn$_9$O$_{19}$. When the direction of the applied magnetic field is inverted, the magnetization of RbSb$_3$Mn$_9$O$_{19}$ displays sudden drop/increase, while KSb$_3$Mn$_9$O$_{19}$ does not. The observation of the hysteresis indicates possibilities of static spin freezing in ASb$_3$Mn$_9$O$_{19}$. However, such spin freezing does not exist in the long range, as discussed later in Sections 2.7 & 2.8.

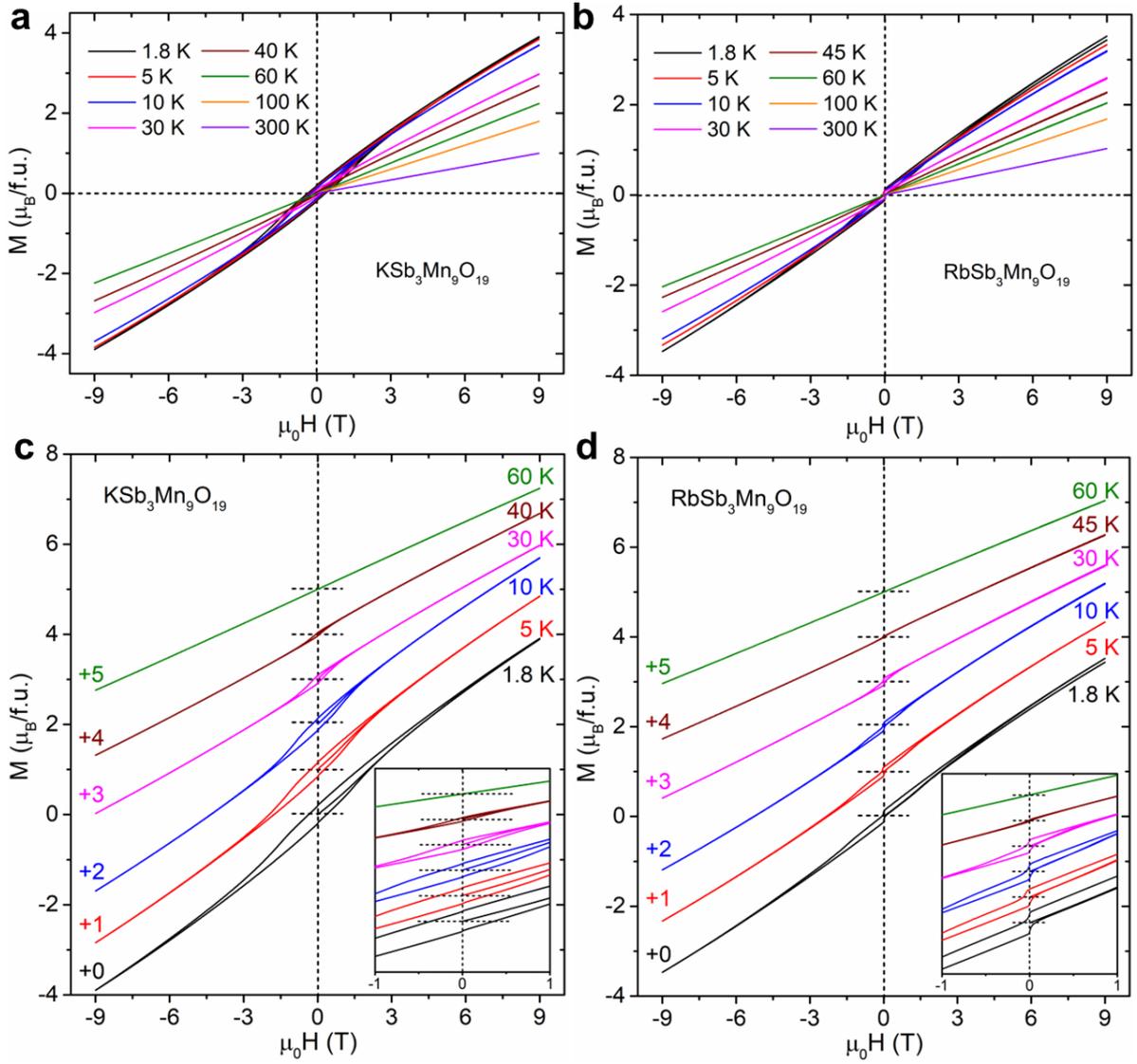

**Figure 4.** Magnetic hysteresis loops of **a.** KSb$_3$Mn$_9$O$_{19}$ and **b.** RbSb$_3$Mn$_9$O$_{19}$ measured under different temperatures. Exploded view of hysteresis loops measured below 60 K between -9 T and 9 T for **c.** KSb$_3$Mn$_9$O$_{19}$ and **d.** RbSb$_3$Mn$_9$O$_{19}$. The loops are offset for clarity. The insets of **c.** and **d.** are the zoom-ins of the main panel between -1 T and 1 T.

**2.5. Single Valency and Oxidation State of Mn:** From the CW fitting results, the fitted $\mu_{eff}$ suggests the Mn$^{3+}$ oxidation state in KSb$_3$Mn$_9$O$_{19}$ since it is close to the spin-only moment of trivalent Mn. However, the fitted $\mu_{eff}$ of RbSb$_3$Mn$_9$O$_{19}$ exhibits a higher value (~5.51 μ$_B$/Mn), which can be attributed to two possible reasons: there are unaccounted contributions from orbital angular momentum, as described in Section 2.4; there is an additional oxidation state, e.g., Mn$^{2+}$, which can result in a larger number of unpaired electrons as well as a bigger $\mu_{eff}$. In order to confirm the number of valency and the oxidation states of Mn, XPS measurements were performed on both samples for the binding energy region of Mn 2$p$ orbitals. As shown in Figures S3a and S3b in SI, an excellent match between the observed curve and fitted curve can be seen. Shirley background and Gaussian peak shape were applied to the fittings. For both compounds, a single pair of 2$p_{3/2}$ and 2$p_{1/2}$ peaks are found, along with a pair of satellite peaks at slightly higher binding energy, as observed in other Mn-based systems,[75,76] implying single valency for Mn in ASb$_3$Mn$_9$O$_{19}$. The binding energy of 2$p_{3/2}$ and 2$p_{1/2}$ peaks in KSb$_3$Mn$_9$O$_{19}$ are 640.61 eV and 652.50 eV, while for RbSb$_3$Mn$_9$O$_{19}$ they are 640.59 eV and 652.43 eV. The similarity of binding energies indicates that Mn in both compounds possesses similar oxidation state, i.e., Mn$^{3+}$ ($S$ = 2). To further prove this, we performed syntheses of ASb$_3$Mn$_9$O$_{19}$ in air by using Mn$_2$O$_3$ as starting material, instead of MnO. The powder XRD pattern with Rietveld fitting of KSb$_3$Mn$_9$O$_{19}$ is shown in Figure S4, which exhibits high purity and great match between the observed crystal structure and observed powder XRD pattern. Considering that Mn$_2$O$_3$ is not likely to be reduced in air under high temperatures, this further proves that Mn in ASb$_3$Mn$_9$O$_{19}$ should adopt oxidation states ≥ 3+. Therefore, the larger fitted $\mu_{eff}$ should originate from unaccounted orbital angular momentum contributions.

**2.6. Heat Capacity:** To investigate whether long-range magnetic order exists in ASb$_3$Mn$_9$O$_{19}$, temperature-dependent heat capacity (C$_p$) measurements were conducted under zero magnetic field from 2 K to 100 K. As shown in the main panels of Figures 5a and 5b, C$_p$ of both materials decrease monotonically and approximately linearly with decreasing temperature. Two small kinks can be seen in the inset of Figure 5a for KSb$_3$Mn$_9$O$_{19}$ at ~38 K and ~45 K, which shows the temperature-dependent C$_p$/T. The two kinks are likely to be relevant to T$_3$ and T$_4$ illustrated in Figure 3c. Kinks at ~10 K, ~26 K and ~43 K can also be observed in the C$_p$/T vs T curve of RbSb$_3$Mn$_9$O$_{19}$, likely correlated with T$_1$, $T_2$ and $T_3$ in Figure 3d. Additionally, no obvious peak/λ-anomaly is seen in the heat capacity of both compounds, suggesting that the associated magnetic entropy is minimal. The typical reasons for a small magnetic entropy associated with a long-range magnetic ordering include 1. The magnetic moments of the species are small; 2. The fraction of magnetic species that participates into the magnetic ordering is low. Given the Mn$^{3+}$ ($S = 2$) from magnetic susceptibility and XPS results, which correspond to a large magnetic moment, we speculate that the reason for the minimal heat capacity peaks is that the fraction of Mn participating in the magnetic ordering in ASb$_3$Mn$_9$O$_{19}$ is limited. This is qualitatively consistent with the short-range magnetic order indicated by the neutron diffraction data of RbSb$_3$Mn$_9$O$_{19}$ below.

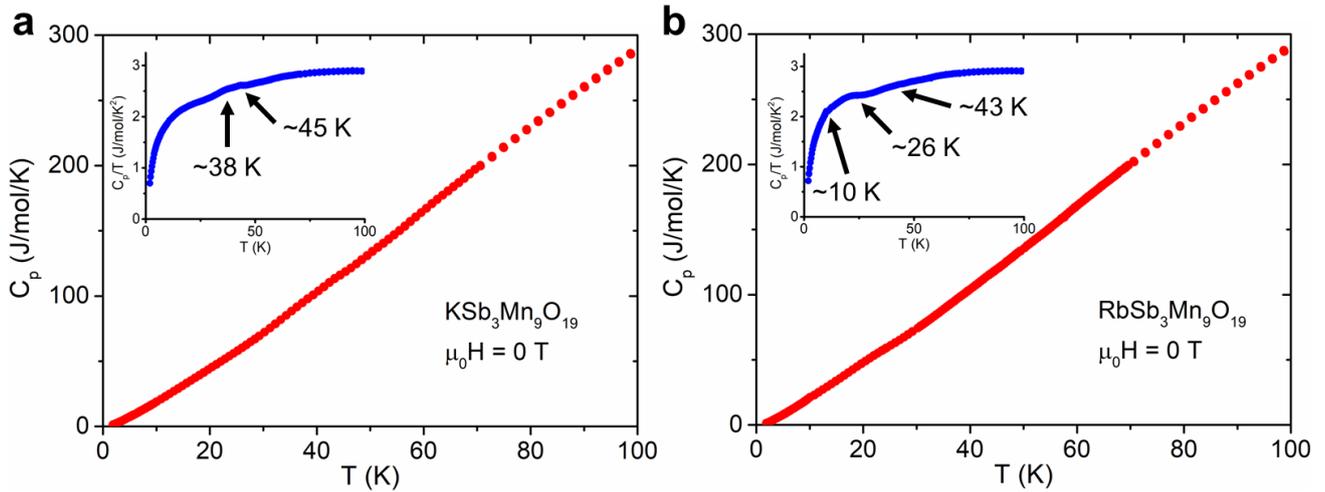

**Figure 5.** Temperature-dependence of heat capacity ($C_p$) between 2 K and 100 K under zero magnetic field of **a.** KSb$_3$Mn$_9$O$_{19}$ and **b.** RbSb$_3$Mn$_9$O$_{19}$. The insets show the temperature-dependence of $C_p/T$.

**2.7. Incommensurate Magnetic Order, Two-Dimensional Magnetic Correlation and Magnetic Frustration:** To better interpret the magnetic behavior and heat capacity of ASb$_3$Mn$_9$O$_{19}$, neutron powder diffraction (NPD) was performed on ASb$_3$Mn$_9$O$_{19}$ powder directly from the reaction crucibles at Oak Ridge National Laboratory. Figures 6a-6f show the NPD patterns of ASb$_3$Mn$_9$O$_{19}$ with Rietveld refinement measured under 300 K, 60 K and 3 K. No additional magnetic peaks are observed when KSb$_3$Mn$_9$O$_{19}$ is cooled to either 60 K or 3 K, indicating the absence of long-range magnetic order. However, at least six magnetic peaks emerge at 3 K for RbSb$_3$Mn$_9$O$_{19}$ below 35°. These sharper magnetic peaks sit on top of broad diffuse scattering, suggesting that not all the Mn sites give rise to these peaks. Meanwhile, the width of the magnetic peaks is larger than that of nuclear peaks, as shown in Figure S5 by comparing the full width at half maximum (FWHM) of the nuclear and magnetic peaks, revealing that $RbSb_3Mn_9O_{19}$ is quasi-long-range ordered. It is noteworthy that the new peaks in $RbSb_3Mn_9O_{19}$ only exist at low Q/2θ, matching the magnetic form factor, while the nuclear structure refinements at 3 K and 60 K do not reveal nuclear structural transition. Therefore, the new peaks indeed originate from symmetry change of the magnetic structure.

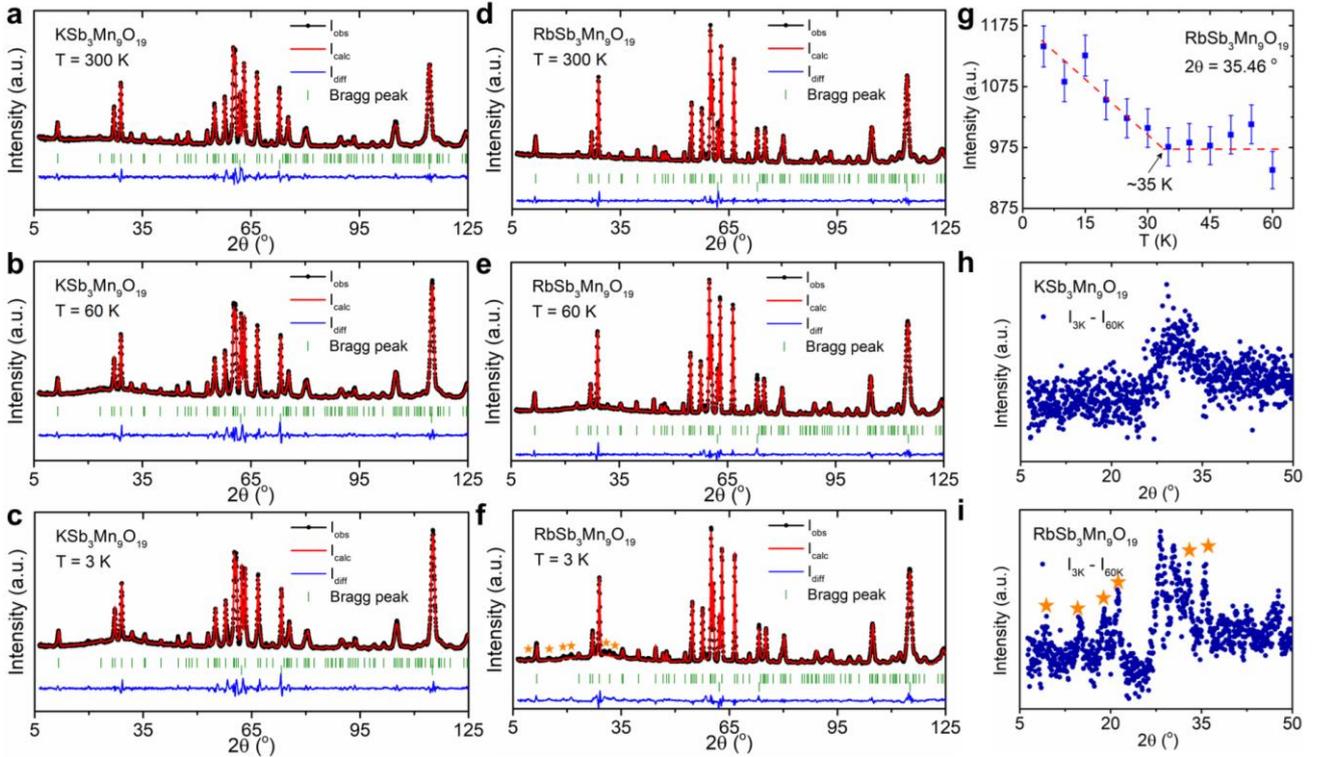

**Figure 6. a-f.** Neutron powder diffraction patterns of $ASb_3Mn_9O_{19}$ under 300 K, 60 K and 3 K with Rietveld fitting. The black solid circle with line and red line represent the observed and calculated patterns, respectively. The blue line stands for the difference between the observed and calculated intensities. The green vertical bars are the Bragg peak positions for $ASb_3Mn_9O_{19}$ (top) and the aluminum can for neutron powder diffraction (bottom). **g.** The evolution of peak intensity at 35.46° for $RbSb_3Mn_9O_{19}$. **h.** and **i.** The temperature subtraction results of both $ASb_3Mn_9O_{19}$ where $I_{3K} - I_{60\ K}$ is shown.

The plot shown in Figure 6g implies a monotonic increase of peak intensity, not the peak area, at 2θ = 35.46° below ~35 K, implying that quasi-long-range magnetic order corresponding to the magnetic peaks in $RbSb_3Mn_9O_{19}$ onsets at ~35 K. The Bragg peak position is selected due to the relatively weaker diffuse scattering at this 2θ such that an obvious increasing intensity can be observed with temperature. Considering that the NPD was measured under zero magnetic field, the onset of quasi-long-range magnetic order in $RbSb_3Mn_9O_{19}$ should correspond to $T_2$ shown in Figure 3d. Efforts were made to index the magnetic propagation vector ($k$-vector) for the magnetic peaks. However, no commensurate $k$-vectors could be found. This implies that the magnetic structure of the quasi-long-range magnetic ordered phase in RbSb$_3$Mn$_9$O$_{19}$ likely possesses an incommensurate magnetic structure. Furthermore, searches of incommensurate $k$-vectors along high-symmetry lines of the Brillouin zone also returned no feasible solutions, suggesting that the incommensurate $k$-vector is of low symmetry. Given that there are also three distinct Mn sites, it is very difficult to figure out the incommensurate $k$-vector with just the powder diffraction patterns.

Other than the incommensurate magnetic order in RbSb$_3$Mn$_9$O$_{19}$, short-range spin order exists in both KSb$_3$Mn$_9$O$_{19}$ and RbSb$_3$Mn$_9$O$_{19}$. Figures 6h and 6i exhibit the subtraction of NPD patterns of 300 K from that of 3 K ($I_{3K} - I_{60K}$), while Figure S6 in the SI presents $I_{3K} - I_{300K}$ and $I_{60K} - I_{300K}$ to show the evolution of the NPD patterns. The minor data points scattered vertically from the majority is due to the thermal expansion of the nuclear peaks, while the magnetic peaks in RbSb$_3$Mn$_9$O$_{19}$ are marked with stars. Note that the subtraction of the high-temperature data from the low-temperature one removes the temperature-independent instrument background, and the resulting patterns are intrinsic to ASb$_3$Mn$_9$O$_{19}$. Interestingly, strong diffuse scattering can be seen in both compounds, indicated by the broad peak centered at ~29°. The broad peak indicates that spin-spin correlations are local rather than long-range. The strong magnetic diffuse scattering observed in NPD usually originates from two-dimensional short-range magnetic correlations. Similar behavior was reported in two-dimensional short-range ordered frustrated spinels, such as Li$_2$Mn$_2$O$_4$.[77–79] Considering the lack of λ-anomaly in heat capacity measurements, as well as the puckered honeycomb, Kagome and triangular sublattices of Mn in ASb$_3$Mn$_9$O$_{19}$, it is likely that magnetic frustration exists in ASb$_3$Mn$_9$O$_{19}$. If taking T$_4$ in Figures 3c and 3d as the first magnetic anomaly temperature, the frustration factors are 3.4 and 3.8 for KSb$_3$Mn$_9$O$_{19}$ and RbSb$_3$Mn$_9$O$_{19}$, respectively. The synthesis of single crystals is indeed needed for further investigations on the magnetism in ASb$_3$Mn$_9$O$_{19}$.

**2.8. Reverse Monte Carlo Refinement:** To determine the nature of the short-range spin correlations, reverse Monte Carlo (RMC) refinement as implemented in the Spinvert program[80] was employed to fit the magnetic diffuse scattering. Details of the data processing and refinement parameters are given in Experimental Details. Figures 7a, 7b, and 7c show the experimental diffuse scattering data with 300 K data subtracted, and the RMC fits for KSb$_3$Mn$_9$O$_{19}$ at 3 K, KSb$_3$Mn$_9$O$_{19}$ at 60 K, and RbSb$_3$Mn$_9$O$_{19}$ at 60 K, respectively (we note the data for RbSb$_3$Mn$_9$O$_{19}$ at 3 K could not be modeled using this approach due to the presence of sharp incommensurate peaks). At 60 K, the diffuse scattering data is similar for A = K and Rb. At 3 K, the diffuse scattering data for A = K shows sharper features compared to 60 K, indicating an increase in the magnetic correlation length on cooling the sample. To quantify the magnetic correlations, the spin-spin correlation function <**S**(0).**S**(r)> was calculated and is shown in Figures 7d, 7e, and 7f for KSb$_3$Mn$_9$O$_{19}$ at 3 K, KSb$_3$Mn$_9$O$_{19}$ at 60 K, and RbSb$_3$Mn$_9$O$_{19}$ at 60 K, respectively. This quantity is equivalent to the magnetic pair distribution function for magnetically-isotropic systems; it is equal to +1 if spin pairs separated by distance *r* are perfectly aligned on average, and to –1 if perfectly anti-aligned. It reveals that antiferromagnetic correlations persist at 60 K for both A = K and Rb, with similar magnitudes in both compounds. The strongest correlation is antiferromagnetic and occurs in a short-range manner (≲ 4 Å) between Mn2-Mn2, i.e., within the Kagome layers, which is consistent with its shortest Mn-Mn distance. Other than Mn2-Mn2, weaker magnetic correlation can be observed between Mn1-Mn1, Mn1-Mn2 and Mn2-Mn3 pairs; i.e., primarily within Kagome-honeycomb bilayers, suggesting mainly two-dimensional correlations. Given that similar magnetic diffuse scattering was observed in other two-dimensional short-range ordered frustrated magnets,[77-79] we can conclude that two-dimensional magnetic correlation indeed exists in $ASb_3Mn_9O_{19}$. At 3 K for A = Rb, the spin correlations show similar trends to at 60 K, but with larger near-neighbor antiferromagnetic correlations and smaller but significant ferromagnetic correlations at ~5.24 Å. The Spinvert fitting confirms the lack of long-range magnetic order in $KSb_3Mn_9O_{19}$ down to a temperature of 3 K, and the existence of antiferromagnetic short-range order in both compounds at 60 K. Together with the values of the frustration parameter mentioned in Section 2.7, this suggests appreciable magnetic frustration.

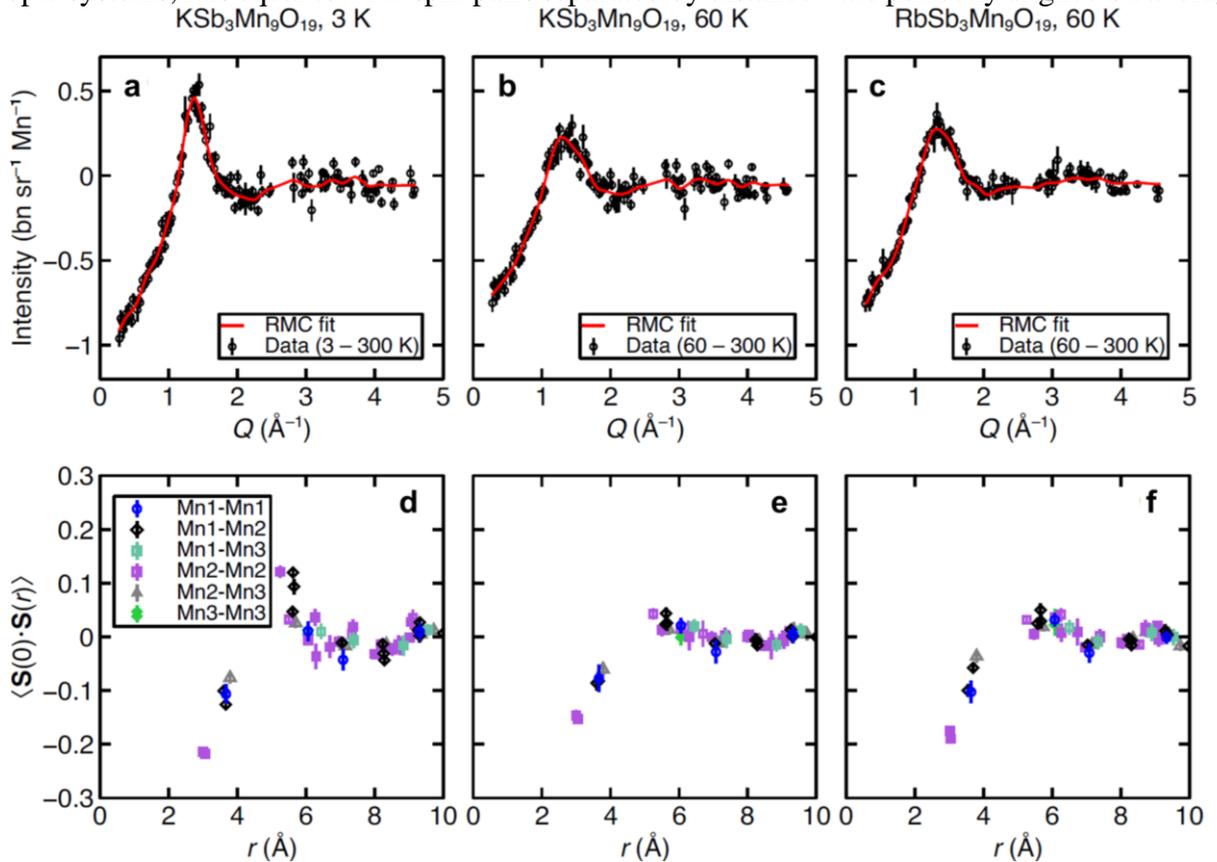

**Figure 7.** Magnetic diffuse scattering in the paramagnetic phase fitted by the Spinvert program of **a.** KSb$_3$Mn$_9$O$_{19}$ at 3 K, **b.** KSb$_3$Mn$_9$O$_{19}$ at 60 K, and **c.** RbSb$_3$Mn$_9$O$_{19}$ at 60 K. **d.**, **e.** and **f.** display the radial dependence of the spin-pair correlation function. The signs indicate that the spins are parallel (+) or antiparallel (-). The values stand for the collinearity between spins.

**2.9. Electrical Resistivity:** Temperature-dependent electrical resistivity of $ASb_3Mn_9O_{19}$ under zero magnetic field were illustrated in Figures 8a and 8b. Both materials exhibit a similar trend of temperature dependence where the resistivity increases monotonically with decreasing temperature, indicating a semiconducting/insulating nature. The signal is out of the instrument's measurable range below the lowest measured temperatures (≲ 350 K). To obtain the electronic bandgap ($E_g$), the Arrhenius equation

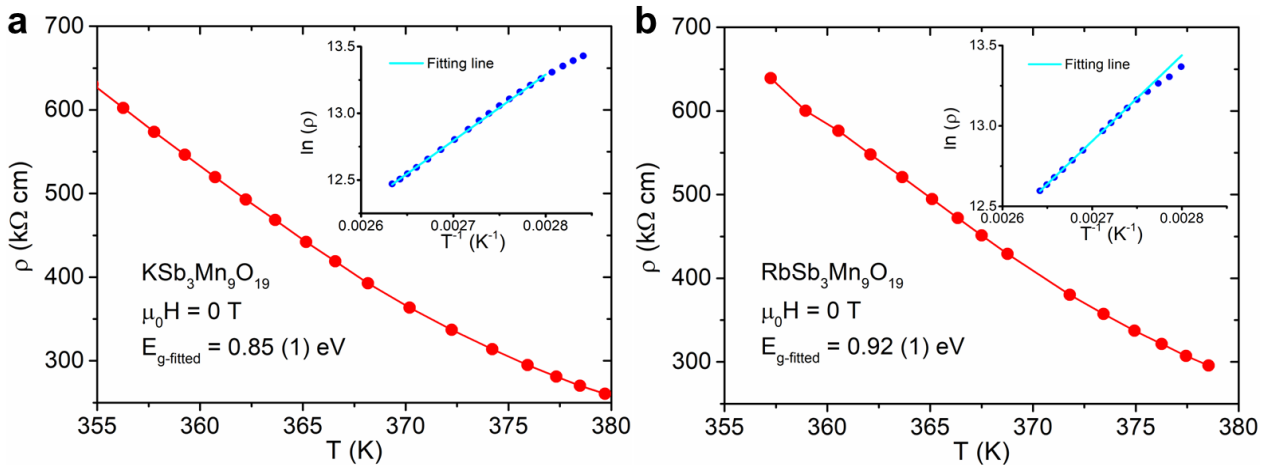

**Figure 8.** Temperature-dependent electrical resistivity of **a.** $KSb_3Mn_9O_{19}$ and **b.** $RbSb_3Mn_9O_{19}$ from ~355 K to ~380 K under no external magnetic field. The insets show the Arrhenius fitting on the $\ln\rho$ vs $T^{-1}$ plots.

$$\rho = \rho_0 e^{E_g/2k_BT}$$

where $\rho_0$ is the pre-exponential term and is a constant, and $k_B$ is Boltzmann's constant, was employed to fit the data, as shown in the insets of Figures 8a and 8b. The fitting was applied only to the highest temperatures where a linear relation between $ln\rho$ and $1/T$ can be found. The fitted bandgap for $KSb_3Mn_9O_{19}$ ($E_{g-K}$) is 0.85 (1) eV and $E_{g-Rb}$ is 0.92 (1) eV, indicating that both $KSb_3Mn_9O_{19}$ and $RbSb_3Mn_9O_{19}$ are semiconductors.

## *3. Conclusion*

Herein, we present the discovery, synthesis and characterization of magnetic properties of the first Mn-based magnetoplumbite phase (M-type hexaferrites), $ASb_3Mn_9O_{19}$ (A = K or Rb). Single-phase polycrystalline samples were synthesized using the high-temperature solid-state method. Single-crystal XRD revealed a crystal structure of both compounds, isostructural to magnetoplumbites. DC magnetic susceptibility and magnetization indicated strong antiferromagnetic coupling under high temperatures and $Mn^{3+}$ ($S = 2$) for $ASb_3Mn_9O_{19}$ as well as multiple magnetic features under low temperatures. However, the temperature-dependent heat capacity did not exhibit any prominent λ-anomaly. After performing neutron powder diffraction, we concluded that $KSb_3Mn_9O_{19}$ is not long-range magnetically ordered within the measured temperature range, while an incommensurate magnetic structure can be found in $RbSb_3Mn_9O_{19}$. Moreover, both compounds show short-range magnetic order, potentially originating from magnetic frustration, as evidenced by strong diffuse scattering in the neutron powder diffraction data and absence of prominent λ-anomaly in the heat capacity. Therefore, the discovery of the first Mn-based magnetoplumbite phase provides a promising platform for studying the intertwining of different geometrically frustrated sublattices of magnetic species. Further investigations on $ASb_3Mn_9O_{19}$, especially on single crystal samples, are necessary to uncover their detailed magnetic structure and identify the origin of magnetic frustration. Additionally, the newly discovered Mn-based magnetoplumbites with $S = 2$ provide an excellent playground for investigating integer-spin-frustrated magnets.

## *4. Experimental Details*

**4.1. Synthesis of Polycrystalline ASb$_3$Mn$_9$O$_{19}$:** Polycrystalline ASb$_3$Mn$_9$O$_{19}$ (A = K or Rb) were prepared using high-temperature solid-state method. Stoichiometric mixture of anhydrous K$_2$CO$_3$ (99%, Thermo Scientific), Rb$_2$CO$_3$ (99.8%, Thermo Scientific), Sb$_2$O$_3$ (99.9%, Thermo Scientific) and MnO (99.99%, Thermo Scientific) with 60% and 40% excess of K$_2$CO$_3$ and Rb$_2$CO$_3$, respectively, were thoroughly ground and placed in alumina crucibles. The excess amount of A$_2$CO$_3$ was used to compensate for their vaporization under high temperatures. The mixture was subsequently heated to 1200 °C and kept for 12 hours followed by air quenching. Both samples were re-annealed at the same temperature overnight several times with intermediate grindings until pure phases were obtained. The total heating time under 1200 °C needed for single-phase products is usually less than 36 hours. The obtained products were black powder and air stable. Small hexagonal crystals (up to ~100×100×10 μm$^3$) can be found within all batches. Polycrystalline KSb$_3$Mn$_9$O$_{19}$ can also be synthesized by replacing MnO with Mn$_2$O$_3$ (98%, Thermo Scientific), which exhibits the same results in powder and single crystal X-ray diffraction measurements. However, synthesis of polycrystalline RbSb$_3$Mn$_9$O$_{19}$ using Mn$_2$O$_3$ only yielded minor target compound along with impurities such as Mn$_3$O$_4$ and Mn$_2$Sb$_2$O$_7$. The failure of synthesis may be due to the higher reactivity of Rb$_2$CO$_3$ as well as the lower reactivity of Mn$_2$O$_3$ such that Rb$_2$CO$_3$ vaporizes before the reaction can start or Mn$_2$O$_3$ was oxidized prior to the reaction temperature. Therefore, in this paper, all the samples used for properties characterizations were made from MnO.

**4.2. Single Crystal and Powder X-Ray Diffraction (XRD):** Multiple crystals (~50×50×10 μm$^3$) of both KSb$_3$Mn$_9$O$_{19}$ and RbSb$_3$Mn$_9$O$_{19}$ were picked for single crystal XRD for crystal structure and chemical composition determination. A Bruker D8 QUEST ECO diffractometer equipped with APEX4 software and Mo radiation ($\lambda_{K\alpha}$= 0.71073 Å) was employed for single crystal XRD measurements at room temperature. Crystals were soaked in glycerol and then mounted using a Kapton loop. The Bruker SMART software was utilized for data acquisition while the corrections for Lorentz and polarization effects were included. Numerical absorption correction was made through a crystal-face-indexing method using *XPREP*. The direct method and full-matrix least-squares on F$^2$ procedure within the SHELXTL package were employed to solve the crystal structure.[81,82]

Powder XRD measurements were utilized to determine the phase purity for the polycrystalline samples. A Bruker D2 PHASER with Cu Kα radiation and a LynxEye-XE detector was employed.

The resulting powder XRD patterns were fitted by the Rietveld method using Fullprof Suite.[83] Crystal structures determined by single crystal XRD were used to obtain the calculated powder XRD patterns and to fit observed patterns.

**4.3. Physical Property Measurement:** The DC magnetic susceptibility ($\chi$), defined as $\chi = M/H$ where M is the observed magnetic moment and H is the applied magnetic field, was measured in a Quantum Design physical property measurement system (PPMS) Dynacool (1.8- 300 K, 0- 9 T) equipped with an ACMS II function from 2 to 300 K under various applied magnetic fields. Field-dependent magnetization data was collected at multiple temperatures with applied magnetic fields ranging from -9 T to 9 T. The resistivity measurements were carried out in the PPMS using the four-probe method between 1.8 K to 300 K. Platinum wires were attached to the samples by silver epoxy to ensure ohmic contact. Heat capacity was measured using a standard relaxation method in the PPMS, with a $^3$He function for data below 1.8 K. All the physical properties measurements were conducted on powder samples (for magnetic measurements) or annealed pelletized samples (for resistivity and heat capacity measurements). For the annealed pelletized samples, they were pressed under a hydraulic press at ~ 20 MPa for 5 minutes before being annealed under 1200 °C for 10 hours. The resulting annealed pellets were utilized for resistivity and heat capacity measurements.

**4.4. Scanning Electron Microscopy with Energy-Dispersive Spectroscopy (SEM-EDS):** Compositional analysis was performed via scanning electron microscopy (SEM) with energy-dispersive spectroscopy (EDS). A Zeiss Sigma 500 VP SEM with Oxford Aztec X-EDS was used with an electron beam energy of 20 kV.

**4.5. X-ray Photoelectron Spectroscopy (XPS):** X-ray photoelectron spectroscopy (XPS) was performed in a Thermo Scientific™ K-AlphaPlus™ instrument equipped with monochromatic Al K$_\alpha$ radiation (1486.7 eV) as the excitation source. The X-ray analysis area for measurement was set at 200 × 400 μm (ellipse shape) and a flood gun was used for charge compensation. The pass energy was 200 eV for the wide (survey) spectra and 50 eV for the high-resolution regions (narrow spectra). The base pressure of the analysis chamber was less than $\sim 1 \times 10^{-9}$ mbar. The analysis chamber pressure was at $1 \times 10^{-7}$ mbar during data acquisition. Data were collected and processed using the Thermo Scientific Avantage XPS software package. Pelletized polycrystalline samples used for XPS measurements were treated the same way as samples used for resistivity and heat capacity measurements.

**4.6. Neutron Powder Diffraction:** Neutron powder diffraction measurements were carried out on the HB-2A powder diffractometer at the High Flux Isotope Reactor (HFIR), Oak Ridge National Laboratory (ORNL).[84,85] Constant wavelength measurements were performed at 2.41 Å from the Ge(113) monochromator reflection. A pyrolytic graphite (PG) filter was placed before the sample to remove higher order reflections. The pre-mono, pre-sample and pre-detector collimation was open-open-12'. The samples were contained in a 6 mm diameter aluminum can and cooled in a closed-cycle refrigerator in the temperature range 3 K - 300 K. The diffraction pattern was collected by scanning a 120º bank of 44 3He detectors in 0.05º steps to give 2Θ coverage from 5º to 130º.

**4.7. Reverse Monte Carlo Refinements:** Neutron powder diffraction data were first placed in absolute intensity units (bn/sr/Mn) by normalization to the nuclear Bragg profile. To remove background and isolate the magnetic scattering signal, data collected at 300 K were subtracted from the data collected at lower temperatures (3 K or 60 K), and data points were excluded where the calculated nuclear Bragg signal exceeded a threshold value. Data were binned in intervals of 0.02 Å$^{-1}$. The magnetic scattering obtained in this way was used as input data for reverse Monte Carlo refinements using the Spinvert program.[80] Refinements with supercells of 8×8×2 crystallographic unit cells (2304 Mn spins). It was assumed that all spins are $S = 2$ with magnetic moments of 4.90 µ$_B$ per Mn, and the Mn1, Mn2, and Mn3 sites were fully occupied; possible Mn occupancy on the Sb1 site was neglected. Refinements were performed for 100 proposed rotations per spin, and a flat background level was allowed to refine to account for changes in the background level between 60 and 300 K. Good agreement with the data was obtained with these parameters.

CCDC 2403656-2403657 contains the supplementary crystallographic data for this paper. These data can be obtained free of charge from The Cambridge Crystallographic Data Centre via www.ccdc.cam.ac.uk/data_request/cif.

*Supporting Information*

Supporting Information is available from the Wiley Online Library or from the author.

*Acknowledgements*

X.G. deeply appreciates the thoughtful discussion with Prof. Graeme Luke (McMaster University) and Prof. Tai Kong (University of Arizona). J.C., G.A. and X.G. thank the support from the startup fund from the University of Pittsburgh. This research used resources at the High Flux Isotope Reactor, a DOE Office of Science User Facility operated by the Oak Ridge National Laboratory. The beam time was allocated to HB-2A (POWDER) on proposal number IPTS-34739.1. Work at the Molecular Foundry was supported by the Office of Science, Office of Basic Energy Sciences, of the U.S. Department of Energy under Contract No. DE-AC02-05CH11231.

Received: xxx

Revised: xxx

Published online: xxx

*Supporting Information*

# ASb$_3$Mn$_9$O$_{19}$ (A = K or Rb): New Mn-Based Two-Dimensional Magnetoplumbites with Geometric and Magnetic Frustration

*Jianyi Chen, Stuart Calder, Joseph A. M. Paddison, Gina Angelo, Liana Klivansky, Jian Zhang, Huibo Cao and Xin Gui\**

Jianyi Chen, Gina Angelo, Xin Gui
Department of Chemistry, University of Pittsburgh, Pittsburgh, PA, 15260, USA
E-mail: xig75@pitt.edu
Stuart Calder, Joseph A. M. Paddison, Huibo Cao
Neutron Scattering Division, Oak Ridge National Laboratory, Oak Ridge, TN, 37831, USA
Liana Klivansky, Jian Zhang
The Molecular Foundry, Lawrence Berkeley National Laboratory, Berkeley, CA, 94720, USA

**Table of Contents**

**Table S1**. Atomic coordinates and equivalent isotropic displacement parameters for ASb$_3$Mn$_9$O$_{19}$ (A = K or Rb) at 293 (2) K. (U$_{eq}$ is defined as one-third of the trace of the orthogonalized U$_{ij}$ tensor (Å$^2$))

KSb$_{2.82(1)}$Mn$_{9.18(1)}$O$_{19}$:

| Atom | Wyck. | Occ. | x | y | Z | U$_{eq}$ |
|---|---|---|---|---|---|---|
| Sb1 | 4f | 1 | 1/3 | 2/3 | 0.18971 (2) | 0.0100 (1) |
| Sb2 | 2a | 0.82 (1) | 0 | 0 | 0 | 0.0079 (3) |
| Mn4 | 2a | 0.18 (1) | 0 | 0 | 0 | 0.0079 (3) |
| Mn1 | 4f | 1 | 1/3 | 2/3 | 0.02130 (6) | 0.0135 (3) |
| Mn2 | 12k | 1 | 0.83341 (8) | 0.16659 (8) | 0.10877 (3) | 0.0105 (2) |
| Mn3 | 2b | 1 | 0 | 0 | ¼ | 0.0201 (5) |
| K1 | 2d | 1 | 2/3 | 1/3 | ¼ | 0.0181 (5) |
| O1 | 12k | 1 | 0.1488 (4) | 0.8512 (4) | 0.0508 (2) | 0.0181 (8) |
| O2 | 12k | 1 | 0.4949 (5) | 0.5051 (5) | 0.1508 (2) | 0.0149 (7) |
| O3 | 6h | 1 | 0.1943 (5) | 0.3885 (11) | ¼ | 0.012 (1) |
| O4 | 4f | 1 | 2/3 | 1/3 | 0.0670 (3) | 0.020 (2) |
| O5 | 4e | 1 | 0 | 0 | 0.1480 (3) | 0.025 (2) |

RbSb$_{2.93(1)}$Mn$_9$O$_{19}$:

| Atom | Wyck. | Occ. | x | y | z | U$_{eq}$ |
|---|---|---|---|---|---|---|
| Sb1 | 4f | 1 | 1/3 | 2/3 | 0.18960 (2) | 0.0075 (1) |
| Sb2 | 2a | 0.931 (5) | 0 | 0 | 0 | 0.0065 (2) |
| Mn1 | 4f | 1 | 1/3 | 2/3 | 0.02146 (6) | 0.0125 (3) |
| Mn2 | 12k | 1 | 0.83332 (8) | 0.16668 (8) | 0.10868 (3) | 0.0089 (2) |
| Mn3 | 2b | 1 | 0 | 0 | ¼ | 0.0196 (4) |
| Rb1 | 2d | 1 | 2/3 | 1/3 | ¼ | 0.0124 (2) |
| O1 | 12k | 1 | 0.1485 (4) | 0.8515 (4) | 0.0503 (2) | 0.0170 (7) |
| O2 | 12k | 1 | 0.4941 (4) | 0.5059 (4) | 0.1503 (1) | 0.0114 (6) |
| O3 | 6h | 1 | 0.1940 (5) | 0.3879 (10) | ¼ | 0.0091 (8) |
| O4 | 4f | 1 | 2/3 | 1/3 | 0.0665 (3) | 0.017 (1) |
| O4 | 4e | 1 | 0 | 0 | 0.1477 (3) | 0.029 (2) |

**Table S2.** Anisotropic thermal displacement parameters for $ASb_3Mn_9O_{19}$ (A = K or Rb) at 293 (2) K.

$KSb_{2.82(1)}Mn_{9.18(1)}O_{19}$:

| Atom | U11 | U22 | U33 | U12 | U13 | U23 |
|---|---|---|---|---|---|---|
| Sb1 | 0.0117 (2) | 0.0117 (2) | 0.0066 (2) | 0.0058 (1) | 0 | 0 |
| Sb2 | 0.0091 (3) | 0.0091 (3) | 0.0054 (4) | 0.0046 (2) | 0 | 0 |
| Mn4 | 0.0091 (3) | 0.0091 (3) | 0.0053 (4) | 0.0046 (2) | 0 | 0 |
| Mn1 | 0.0139 (4) | 0.0139 (4) | 0.0127 (6) | 0.0069 (2) | 0 | 0 |
| Mn2 | 0.0115 (3) | 0.0115 (3) | 0.0105 (3) | 0.0074 (3) | -0.0001 (1) | 0.0001 (1) |
| Mn3 | 0.0099 (5) | 0.0099 (5) | 0.0404 (13) | 0.0050 (3) | 0 | 0 |
| K1 | 0.0188 (8) | 0.0188 (8) | 0.017 (1) | 0.0094 (4) | 0 | 0 |

$RbSb_{2.93(1)}Mn_9O_{19}$:

| Atom | U11 | U22 | U33 | U12 | U13 | U23 |
|---|---|---|---|---|---|---|
| Sb1 | 0.0084 (2) | 0.0084 (2) | 0.0057 (2) | 0.0042 (1) | 0 | 0 |
| Sb2 | 0.0070 (3) | 0.0070 (3) | 0.0054 (3) | 0.0035 (1) | 0 | 0 |
| Mn1 | 0.0130 (4) | 0.0130 (4) | 0.0114 (5) | 0.0065 (2) | 0 | 0 |
| Mn2 | 0.0098 (2) | 0.0098 (2) | 0.0096 (3) | 0.0067 (3) | -0.0003 (1) | 0.0003 (1) |
| Mn3 | 0.0069 (5) | 0.0069 (5) | 0.0449 (13) | 0.0035 (3) | 0 | 0 |
| Rb1 | 0.0126 (3) | 0.0126 (3) | 0.0119 (4) | 0.0063 (2) | 0 | 0 |
| O5 | 0.038 (3) | 0.038 (3) | 0.010 (3) | 0.019 (2) | 0 | 0 |

**Table S3.** Comparison of single crystal structure refinement parameters for KSb$_3$Mn$_9$O$_{19}$ with Sb2 site mixed with Mn atom and constrained to be 100% occupied.

| Refined Formula | KSb$_{2.82(1)}$Mn$_{9.18(1)}$O$_{19}$ | KSb$_3$Mn$_9$O$_{19}$ |
|---|---|---|
| Temperature (K) | 293 (2) | 293 (2) |
| F.W. (g/mol) | 1190.45 | 1202.81 |
| Space group; Z | *P* 6$_3$/mmc; 2 | *P* 6$_3$/*mmc*; 2 |
| *a*(Å) | 6.0606 (1) | 6.0606 (1) |
| *c*(Å) | 23.853 (1) | 23.853 (1) |
| V (Å$^3$) | 758.76 (5) | 758.76 (5) |
| θ range (º) | 3.416-33.191 | 3.416-33.191 |
| No. reflections; $R_{int}$ | 21526; 0.0631 | 21526; 0.0631 |
| No. independent reflections | 620 | 620 |
| No. parameters | 32 | 32 |
| $R_1$: $\omega R_2$ (*I*>2δ(*I*)) | 0.0304; 0.0657 | 0.0380; 0.0908 |
| Goodness of fit | 1.199 | 1.178 |
| Diffraction peak and hole (e$^-$/ Å$^3$) | 1.308; -1.268 | 1.422; -3.665 |

**Table S4.** Comparison of single crystal structure refinement parameters for RbSb$_3$Mn$_9$O$_{19}$ with Sb2 site relaxed and constrained to be 100% occupied.

| Refined Formula | RbSb$_{2.93(1)}$Mn$_9$O$_{19}$ | RbSb$_3$Mn$_9$O$_{19}$ |
|---|---|---|
| Temperature (K) | 293 (2) | 293 (2) |
| F.W. (g/mol) | 1240.66 | 1249.81 |
| Space group; Z | $P6_3/mmc$; 2 | $R\bar{3}m$; 3 |
| $a$(Å) | 6.0818 (7) | 6.0818 (7) |
| $c$(Å) | 23.905 (4) | 23.905 (4) |
| V (Å$^3$) | 765.8 (2) | 765.8 (2) |
| θ range (°) | 3.409-34.343 | 3.409-34.343 |
| No. reflections; $R_{int}$ | 21395; 0.0401 | 21395; 0.0401 |
| No. independent reflections | 681 | 681 |
| No. parameters | 34 | 32 |
| $R_1$: $\omega R_2$ ($I>2\delta(I)$) | 0.0291; 0.0594 | 0.0343; 0.0766 |
| Goodness of fit | 1.387 | 1.325 |
| Diffraction peak and hole (e$^-$/ Å$^3$) | 1.174; -0.892 | 1.315; -3.354 |

**Table S5.** EDS results for $KSb_3Mn_9O_{19}$.

| Spectrum # | K | Sb | Mn | Mn/Sb Ratio |
|---|---|---|---|---|
| Spectrum 1 | 3.1(1) | 9.0(2) | 28.9(2) | 3.2(1) |
| Spectrum 2 | 3.1(1) | 9.2(2) | 29.6(2) | 3.2(1) |
| Spectrum 3 | 3.0(1) | 8.9(2) | 28.4(2) | 3.2(1) |
| Spectrum 4 | 3.0(1) | 9.0(2) | 28.0(2) | 3.1(1) |
| Spectrum 5 | 2.9(1) | 8.8(2) | 27.6(2) | 3.1(1) |
| Spectrum 6 | 2.9(1) | 8.6(2) | 27.9(2) | 3.2(1) |
| Spectrum 7 | 3.0(1) | 8.7(2) | 27.6(2) | 3.2(1) |
| Spectrum 8 | 3.1(1) | 9.1(2) | 28.3(2) | 3.1(1) |
| Spectrum 9 | 2.6(1) | 8.0(2) | 24.6(2) | 3.1(1) |
| Spectrum 10 | 3.0(1) | 8.7(2) | 27.5(2) | 3.2(1) |
| Spectrum 11 | 3.0(1) | 9.3(2) | 31.4(2) | 3.4(1) |
| Spectrum 12 | 3.0(1) | 8.8(2) | 28.9(2) | 3.3(1) |
| Spectrum 13 | 2.8(1) | 8.7(2) | 27.9(2) | 3.2(1) |
| Spectrum 14 | 3.1(1) | 8.5(2) | 27.9(2) | 3.3(1) |
| Spectrum 15 | 3.0(1) | 8.8(2) | 28.1(2) | 3.2(1) |
| Spectrum 16 | 2.9(1) | 8.8(2) | 28.3(2) | 3.2(1) |
| Spectrum 17 | 2.7(1) | 7.7(2) | 24.1(2) | 3.1(1) |
| Spectrum 18 | 3.0(1) | 8.6(2) | 27.7(2) | 3.2(1) |
| Spectrum 19 | 2.9(1) | 8.9(2) | 27.7(2) | 3.1(1) |
| Spectrum 20 | 3.0(1) | 8.6(2) | 27.5(2) | 3.2(1) |
| Spectrum 21 | 3.0(1) | 8.8(2) | 27.4(2) | 3.1(1) |
| Spectrum 22 | 2.8(1) | 8.6(2) | 28.0(2) | 3.3(1) |
| Spectrum 23 | 2.9(1) | 8.5(2) | 27.9(2) | 3.3(1) |
| Spectrum 24 | 3.0(1) | 8.7(2) | 27.1(2) | 3.1(1) |
| Average | 2.95 (10) | 8.73 (19) | 27.85 (0.20) | 3.19 (9) |
| Normalize to K | 1.00 (3) | 2.96 (6) | 9.44 (7) | |

**Table S6.** EDS results for RbSb$_3$Mn$_9$O$_{19}$.

| Spectrum Label | Rb | Sb | Mn | Mn/Sb Ratio |
|---|---|---|---|---|
| Spectrum 1 | 3.0(2) | 8.6(2) | 25.1(2) | 2.9(1) |
| Spectrum 2 | 3.1(2) | 8.7(2) | 25.3(2) | 2.9(1) |
| Spectrum 3 | 2.7(2) | 10.6(2) | 32.3(2) | 3.0(1) |
| Spectrum 4 | 3.1(2) | 9.0(2) | 27.1(2) | 3.0(1) |
| Spectrum 5 | 3.1(2) | 9.0(2) | 26.8(2) | 3.0(1) |
| Spectrum 6 | 3.1(2) | 9.0(2) | 26.9(2) | 3.0(1) |
| Spectrum 7 | 3.0(2) | 9.0(2) | 26.9(2) | 3.0(1) |
| Spectrum 8 | 3.0(2) | 10.5(2) | 32.3(2) | 3.1(1) |
| Spectrum 9 | 3.1(2) | 10.6(2) | 32.7(2) | 3.1(1) |
| Spectrum 10 | 2.9(2) | 10.8(2) | 32.7(2) | 3.0(1) |
| Spectrum 11 | 3.0(2) | 10.6(2) | 32.4(2) | 3.1(1) |
| Spectrum 12 | 3.0(2) | 9.2(2) | 26.9(2) | 2.9(1) |
| Spectrum 13 | 3.1(2) | 9.1(2) | 27.4(2) | 3.0(1) |
| Spectrum 14 | 2.9(2) | 9.1(2) | 27.3(2) | 3.0(1) |
| Spectrum 15 | 3.0(2) | 9.1(2) | 27.4(2) | 3.0(1) |
| Spectrum 16 | 3.1(2) | 9.2(2) | 27.0(2) | 2.9(1) |
| Spectrum 17 | 3.0(2) | 9.1(2) | 27.6(2) | 3.0(1) |
| Spectrum 18 | 3.0(2) | 9.0(2) | 27.6(2) | 3.1(1) |
| Spectrum 19 | 3.0(2) | 9.6(2) | 29.3(2) | 3.1(1) |
| Spectrum 20 | 3.2(2) | 9.5(2) | 27.9(2) | 2.9(1) |
| Spectrum 21 | 3.1(2) | 9.4(2) | 28.0(2) | 3.0(1) |
| Spectrum 22 | 2.7(2) | 9.7(2) | 29.1(2) | 3.0(1) |
| Spectrum 23 | 3.0(2) | 10.4(2) | 30.8(2) | 2.9(1) |
| Average | 3.01 (34) | 9.51 (20) | 28.56 (19) | 3.00 (9) |
| Normalize to Rb | 1.0 (1) | 3.16 (7) | 9.49 (6) | |

**Table S7.** Comparison of Mn/Sb ratio obtained from single crystal X-ray diffraction (SCXRD) and energy-dispersive spectroscopy (EDS).

| Sample name | Situation on Sb2 site | SCXRD | EDS |
|---|---|---|---|
| **KSb$_3$Mn$_9$O$_{19}$** | Mn-Sb mix | 3.26 (2) | 3.19 (9) |
|  | Sb with vacancy | 3.10 (1) |  |
| **KSb$_3$Mn$_9$O$_{19}$** | Mn-Sb mix | 3.18 (2) | 3.00 (9) |
|  | Sb with vacancy | 3.07 (1) |  |

**Figure S1.** Possible Mn-Mn superexchange pathways in ASb$_3$Mn$_9$O$_{19}$.

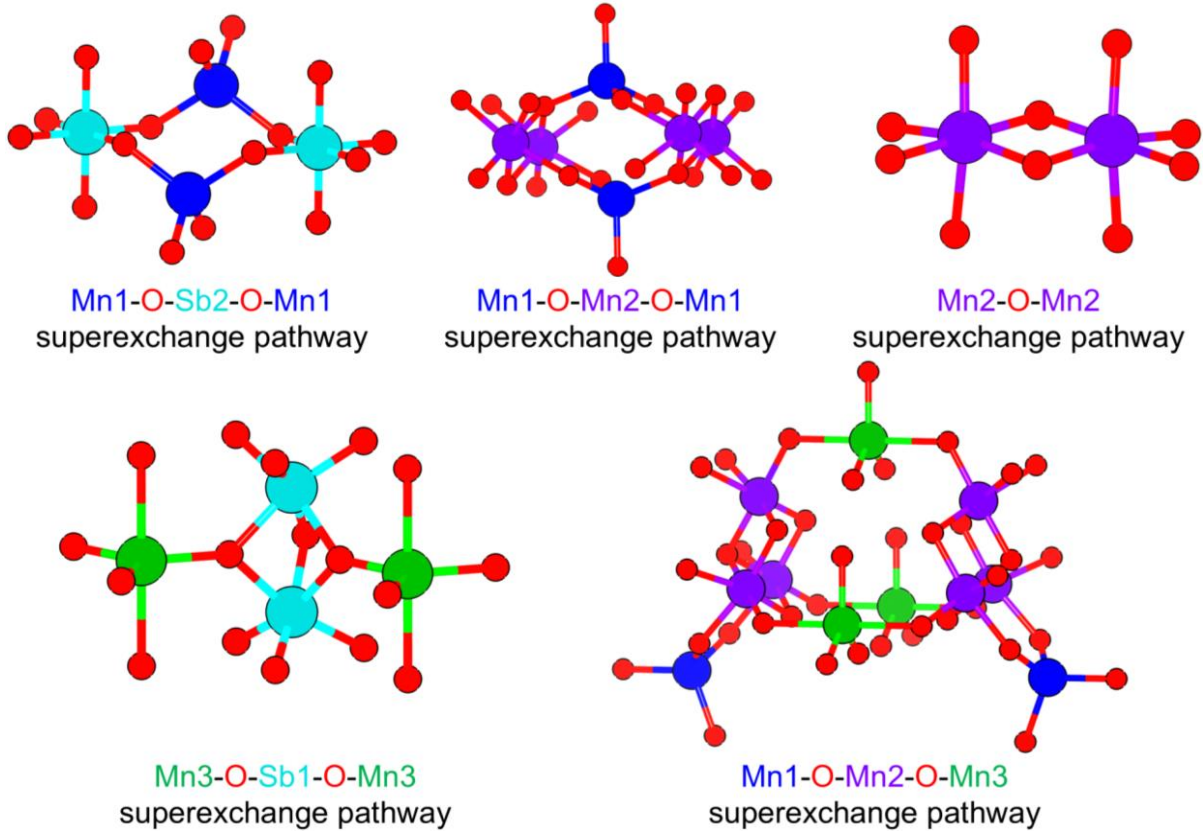

**Figure S2.** Temperature-dependent magnetic susceptibility with both ZFC and FC modes under different magnetic fields of **(a-e)** KSb$_3$Mn$_9$O$_{19}$ and **(f-j)** RbSb$_3$Mn$_9$O$_{19}$.

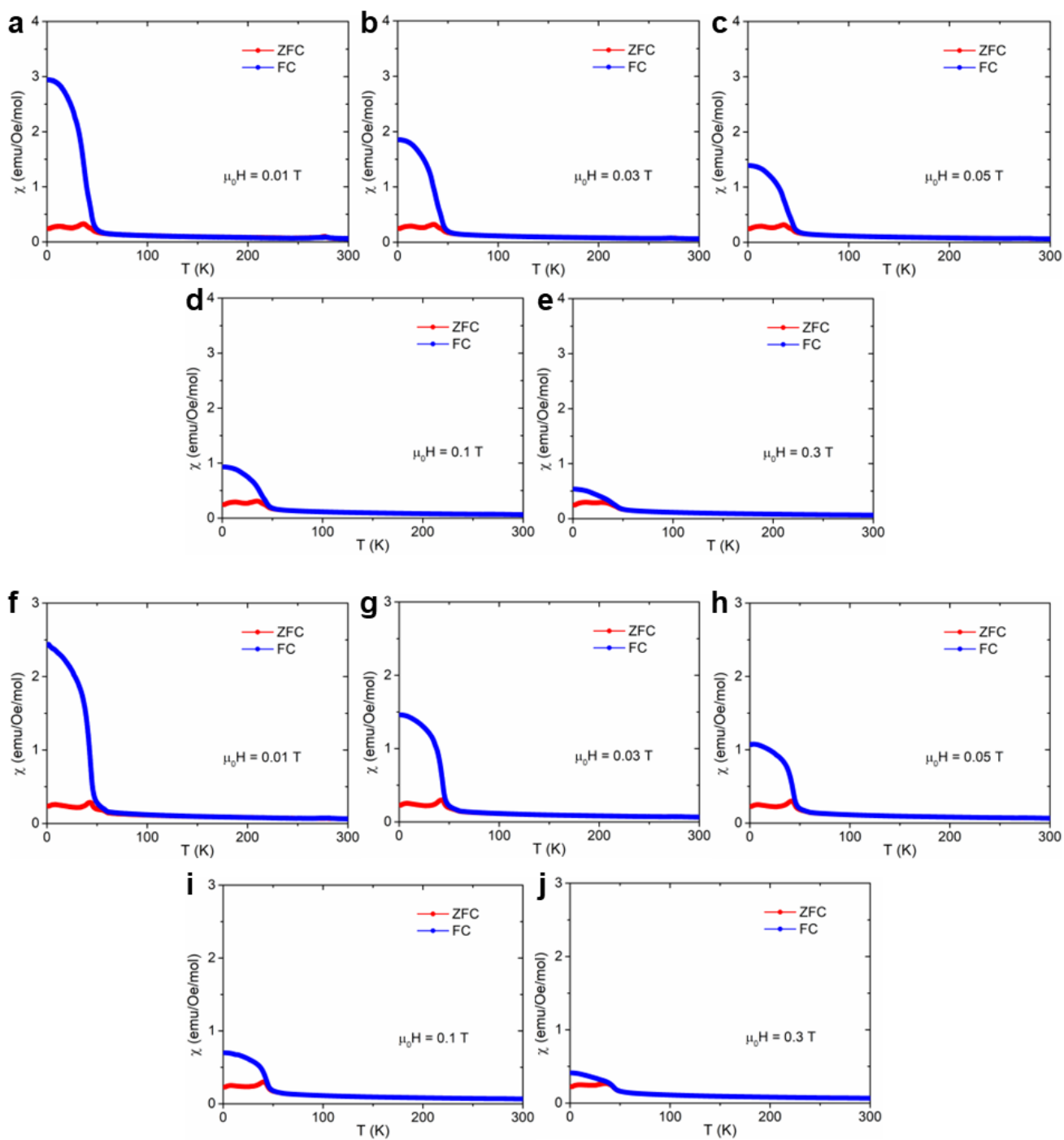

**Figure S3.** XPS spectra of Mn 2*p* orbitals for **a.** $KSb_3Mn_9O_{19}$ and **b.** $RbSb_3Mn_9O_{19}$.

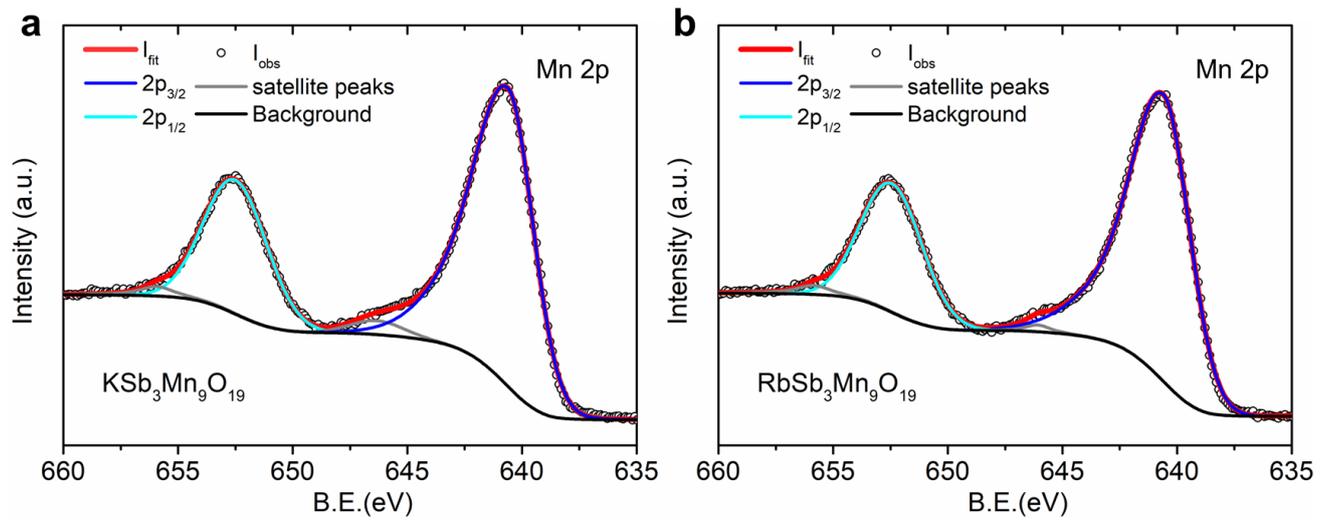

**Figure S4.** Powder X-ray diffraction patterns of $KSb_3Mn_9O_{19}$ with $Mn_2O_3$ as starting material.

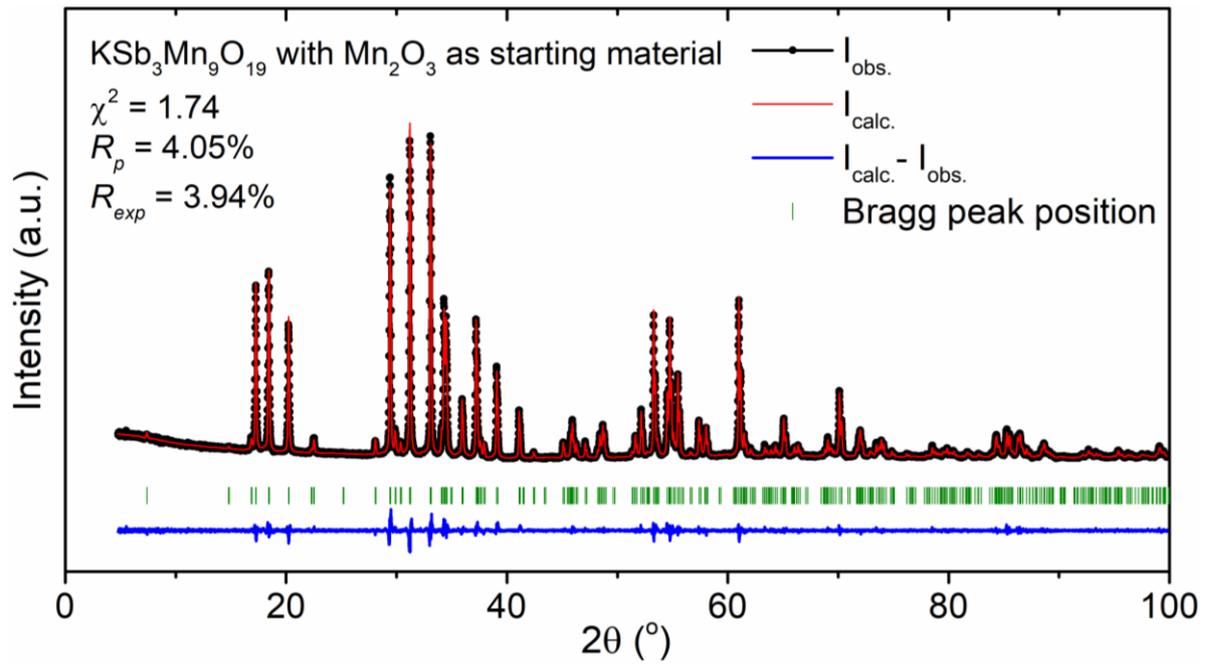

**Figure S5.** The peak width at low 2θ of RbSb$_3$Mn$_9$O$_{19}$. The nuclear peaks and magnetic peaks and their full width at half maximum (FWHM) are marked by blue and orange, respectively.

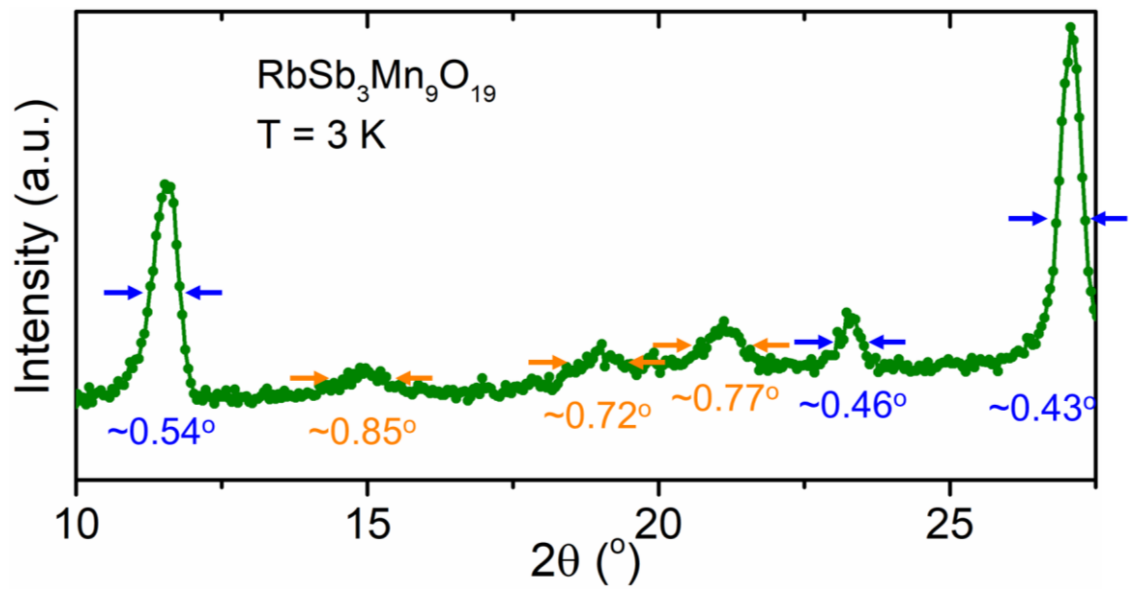

**Figure S6.** Evolution of diffuse scattering in ASb₃Mn₉O₁₉ represented by subtracting the neutron powder diffraction pattern at 300 K from that of 60 K, as well as subtracting the pattern at 300 K from that of 3 K.

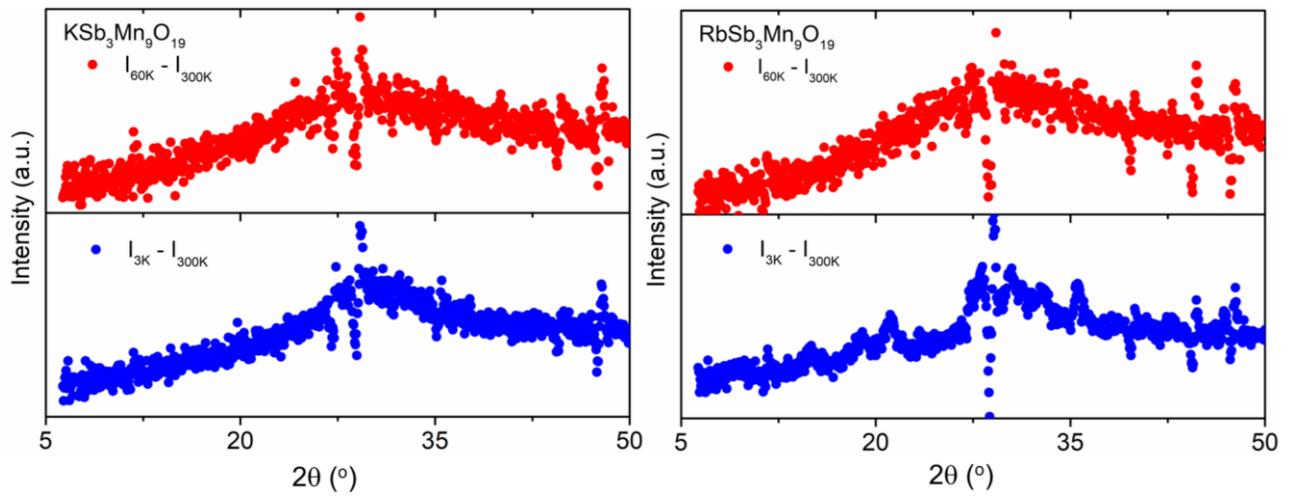